\documentclass[sigconf]{acmart}

\AtBeginDocument{%
  }

\copyrightyear{2026}
\acmYear{2026}
\setcopyright{cc}
\setcctype{by}
\acmConference[CHI '26]{Proceedings of the 2026 CHI Conference on Human Factors in Computing Systems}{April 13--17, 2026}{Barcelona, Spain}
\acmBooktitle{Proceedings of the 2026 CHI Conference on Human Factors in Computing Systems (CHI '26), April 13--17, 2026, Barcelona, Spain}
\acmPrice{}
\acmDOI{10.1145/3772318.3790508}
\acmISBN{979-8-4007-2278-3/2026/04}

\usepackage{enumitem}
\usepackage{tabularx} 
\usepackage{graphicx}   
\usepackage{array}      
\usepackage{booktabs}   
\newcolumntype{C}[1]{>{\centering\arraybackslash}m{#1}}
\newcommand{\secicon}[2][1.2ex]{%
  \texorpdfstring{\raisebox{-0.2ex}{\includegraphics[height=#1]{#2}}\hspace{0.35em}}{}%
}
\newtoggle{revisions}
\iftoggle{revisions} {
\newcommand{\rev}[1]{{\color{blue}{#1}\normalfont}}
}{
  \newcommand {\rev}[1]{#1} 
 }
\begin{document}

\title{Restoration, Exploration and Transformation: How Youth Engage Character.AI Chatbots for Feels, Fun and Finding themselves}

\author{Annabel Blake}
\email{abla1903@uni.sydney.edu.au}
\orcid{1234-5678-9012}
\affiliation{%
  \institution{The University of Sydney}
  \city{Sydney}
  \state{NSW}
  \country{Australia}
}
\author{Marcus Carter}
\email{marcus.carter@sydney.edu.au}
\orcid{XXX}
\affiliation{%
  \institution{The University of Sydney}
  \city{Sydney}
  \state{NSW}
  \country{Australia}
}
\author{Eduardo Velloso}
\email{eduardo.velloso@sydney.edu.au}
\orcid{XXX}
\affiliation{%
  \institution{The University of Sydney}
  \city{Sydney}
  \state{NSW}
  \country{Australia}
}

\renewcommand{\shortauthors}{Blake et al.}

\begin{abstract}
Young people are among the fastest adopters of generative AI, yet research emphasises adult-designed tools and experiments rather than playful, self-directed youth use. We analysed discourse from 4,172 users in Character.AI’s official Discord, finding that the most engaged users were predominantly adolescents (50\% aged 13–17), female or non-binary (61.9\%), with most (59\%) creating their own characters. We contribute (1) a descriptive account of how \rev{highly-engaged youth on Character.AI's Discord} use AI for playful, emotional, and creative practices that push the platform limits; (2) a framework of three engagement intents — Restoration (emotional regulation), Exploration (creative experimentation), and Transformation (identity development); and (3) a taxonomy of seven youth-created character archetypes. Together, these findings reveal how youth invent novel roles for AI, expose critical misalignments between youth use and current AI experiences, and provide frameworks for researchers and practitioners to design youth-centred AI futures.
\end{abstract}



\maketitle

\section{Introduction}

On a Saturday morning in a suburban park, children gather on a hillside playground. Ignoring the carefully designed ladders and slides, they tear down the hill using cardboard scraps, inventing their own paths and rules. This moment of spontaneous, self-directed play reflects a broader truth: young people frequently subvert adult-designed systems to meet their own needs. As generative AI enters the mainstream, we observe a digital parallel. Children and adolescents are not just passive consumers of AI; they actively reshape these systems for creative, emotional, and playful purposes.

The pace of youth AI adoption is explosive: 77\% of U.K. teens aged 13–18 used generative AI in 2024, up from 37\% in 2023  \cite{picton_children_2024}, and over 72\% of U.S. teens have used AI \cite{robb_talk_2025}. Critically, youth are not predominantly using AI for the productivity, formal therapy and educational purposes that dominate research and industry attention. Rather, 40\% of children use AI for ``entertainment and playing around''~\cite{hashem_understanding_2025} and 63\% of 8-15 year olds claim to have used AI `for fun', ranking above work and study~\cite{Ofcom2024OnlineNation}.  Character.AI exemplifies this trend, explicitly marketing itself as an ``AI entertainment company'' with over 20 million monthly active users engaging with 10+ million novel AI characters \cite{Shazeer2024, WhatPluginCharacterAI}. The company recently disclosed that over 50\% of its users are Gen Z or Gen Alpha, spending on average 75 minutes on the site daily, with \rev{anecdotal reports of 14-year-olds using Character.AI for over 7 hours in a single day \cite{yu2025understanding}}.\footnote{\url{https://www.wired.com/story/character-ai-ceo-chatbots-entertainment/}} 

Despite this popularity, we know little about youth’s experiences \rev{``playing around''} with Character.AI \cite{hashem_understanding_2025}. Today, academic work focuses on prescribed educational or therapeutic uses in controlled settings rather than examining playful, self-directed use~\cite{wang_informing_2022, mansoor_conversational_2025} and most industry tools are designed through adult lenses, misaligned with the values, needs and behaviours of youth~\cite{zhang_dark_2025, kurian_developmentally_2025, kirk_why_2025}. As a result, the field risks designing AI experiences and safety measures based on imagined rather than actual youth behaviour, further sidelining youth voices. The stakes are not just theoretical: recent lawsuits in 2024/25 have linked the deaths by suicide of two minors to role-play interactions with AI characters, underscoring the urgent need to understand youth intent and unmet needs.\footnote{\url{https://www.prnewswire.com/news-releases/center-for-humane-technology-new-federal-lawsuit-reveals-Character.AI-chatbots-predatory-deceptive-practices-302284593.html}}\footnote{\url{https://www.nytimes.com/2025/08/26/technology/chatgpt-openai-suicide.html?unlocked_article_code=1.hE8.T-3v.bPoDlWD8z5vo&smid=url-share}}.

To address this gap, we analysed discourse from 4,172 users in Character.AI’s global Discord community. \rev{Character.AI is a popular product broadly focused on AI entertainment, and the official Discord community is one of the few accessible spaces where adolescents organically discuss their self-directed AI use with peers}\cite{Shazeer2024}. We explore:
\begin{itemize}
    \item \textbf{RQ1}: What are the demographics of Character.AI's most engaged youth audience?
                \item \textbf{RQ2}: What are the use cases and motivations that youth discuss engaging in? 
                \item \textbf{RQ3}: What are the emerging AI character roles that youth are creating and consuming?
\end{itemize}

Our findings reveal that Character.AI’s most engaged \rev{Discord community} users are primarily adolescents (50\% aged 13–17) and the majority are female or non-binary (61.9\%). An analysis of this cohort's discussions supported three contributions:
\begin{enumerate}
    \item A \textbf{descriptive account} of how young people use \rev{Character.AI} for playful, emotional, and creative practices that often push beyond platform limitations, revealing misalignment between youth use and current AI design.
    \item A \textbf{framework} of three youth engagement patterns (restoration, exploration, and transformation) that operationalises youth motivations and provides practitioners with tools to improve alignment, design decisions, and future research.
    \item A \textbf{taxonomy} of 7 AI character archetypes that youth engage with (Soother, Narrator, Trickster, Icon, Dark Soul, Proxy, Mirror), revealing the need for archetype-specific research \& design approaches.

\end{enumerate}

This research foregrounds youth voices and re-frames the centrality of play to understanding how youth push the boundaries of new technologies in surprising ways. 

\section{Related work}

Understanding how children engage with AI characters requires examining three interconnected challenges in current research: (1) a focus on adult-designed AI interventions that miss emerging youth behaviours, (2) the under-representation of playful AI use, and (3) the deployment of AI entertainment technologies that outpace research understanding.

\subsection{Youth AI research is primarily adult-centric}

Child-AI research frequently excludes authentic youth voices, focusing instead on adult-designed applications of AI to support formal educational and mental health outcomes~\cite{wang_informing_2022, chin_young_2024, han_design_2023, yan_social_2025, tucek_enhancing_2024, dmello_improving_2024, mansoor_conversational_2025}. These studies often show mixed outcomes\cite{APA2025AIAdolescents}: while a few report gains in motivation or language learning \cite{Song2023}, others show limited impact on broader wellbeing measures \cite{mansoor_conversational_2025} or differing impacts on loneliness depending on use patterns~\cite{herbener_are_2025, fang_how_2025}. Risks such as algorithmic bias \cite{weissburg_llms_2025}, cognitive offloading \cite{Jose2025}, AI's `empathy gap' \cite{kurian_ais_2025}, exposure to harmful content \cite{CCDH2024FakeFriend} and social displacement \cite{herbener_are_2025} are acknowledged, but this body of work largely evaluates prescribed use cases or red teaming (systematic stress-testing of AI systems by researchers to identify potential harms). This risks focusing attention on how adults \textit{want} or \textit{imagine} young people to use AI as opposed to how they \textit{actually} use it. 

Study designs tend to reinforce this gap. Youth are typically assigned specific tasks or prototypes \cite{xu_examining_2024}, often due to ethical constraints, and real-world conversational data is often inaccessible or anonymised, obscuring child-specific behaviour \cite{tamkin_clio_2024, phang_investigating_2025}. As a result, self-directed, playful, and emotional youth use surfaces only when things go wrong \footnote{\url{https://www.prnewswire.com/news-releases/center-for-humane-technology-new-federal-lawsuit-reveals-Character.AI-chatbots-predatory-deceptive-practices-302284593.html}}. For example, the young person at the centre of the 2024 Character.AI lawsuit engaged in fantasy roleplay that extended into conversations about deeply personal issues; talking not just \textit{to} a Game of Thrones character, but \textit{as} a character themselves using the platform's persona feature rather than an expertly designed `psychologist' chatbot service.  

Together, it is evidence that the field still has a fundamental gap: while we understand how children respond to adult-designed AI experiences in controlled settings, we know little about their self-directed, recreational AI use, the very contexts where they demonstrate the most engagement. Fittingly, a recent closed-ended survey by Common Sense Media suggests 33\% of teen responders use `AI companions' in a way that is yet to be captured \cite{robb_talk_2025, caldwell_carnegie_2024}. As a result, policy and design recommendations are limited in their understanding, efficacy and applicability; impacting designers who are left without child-centric guidance that goes beyond keeping children safe, to helping them \textit{thrive}~\cite{wang_informing_2022, la_fors_toward_2024, mansfield_social_2025}. 

\subsection{We lack an understanding of playful use of AI}

Whilst 63\% of young people use AI ``for fun''~\cite{Ofcom2024OnlineNation}, and 30\% use companions because they are `entertaining' \cite{hashem_understanding_2025, Ofcom2024OnlineNation, robb_talk_2025}, academic research overwhelmingly focuses on productivity and mental health applications, missing the playful, creative, and improvisational ways young people use technology~\cite{ito_hanging_2019}. Recent analyses of conversational AI usage patterns focus on coding and business applications as the dominating use case~\cite{tamkin_clio_2024}. In educational contexts, Anthropic found that 39\% of conversations focused on creating educational content, with 33.5\% supporting assignment completion\footnote{\url{https://www.anthropic.com/news/anthropic-education-report-how-university-students-use-claude}}. This \rev{research focus reflects} how AI companies typically market their products, as productivity enhancers and learning tools, which may \rev{downplay} emerging playful uses of AI.  

\rev{When platforms advertise entertainment and play as primary use cases, a different research focus is required}~\cite{tamkin_clio_2024}. For example, when a young person logs into Character.AI, they encounter categories for ``games'' and ``novelty'' as well as scenes and games you can play \textit{with} characters. The platform explicitly markets itself as an ``AI entertainment company'' \footnote{\url{https://techcrunch.com/2025/01/17/ai-startup-character-ai-tests-games-on-the-web/}}, fundamentally different from productivity-focused AI assistants. Given that 85\% of US teens play video games \footnote{\url{https://www.pewresearch.org/internet/2024/05/09/teens-and-video-games-today/?utm_source=chatgpt.com}}, platforms like Character.AI may naturally invite the playful, recreational engagement that children report but that remains largely absent from academic research. 

Not only is play missing from research, but there is evidence that playful youth behaviour is misclassified. For instance, while Anthropic's analysis did identify Dungeons \& Dragons roleplay clusters, their classifiers frequently mislabeled these conversations as harmful content~\cite{tamkin_clio_2024}. This is not an outlier use case, with recent research reporting that up to 19\% of young people used generative AI to write a story, including young people who said they typically did not enjoy writing stories \cite{picton_children_2024}. Recent research classifying harms appears to mix in possible examples of playful trolling, inferring distress from users' responses that may be better explained by youth playfulness \cite{zhang_dark_2025, yu2025understanding}. In this, we see a risk that imaginative, creative, and recreational use by youth is underestimated or treated as problematic and unethical. 

Existing child-AI frameworks used in the academic fields of media, communications, and psychology may be insufficient for understanding playful youth engagement with AI. Parasocial relationship theories \rev{narrowly define} youth behaviour as seeing AI as ``best friends'', whereas emerging findings show that youth see it as a third category, as a game, interactive diary, coping or social support tool rather than a friend~\cite{hartmann_horton_2011, bond_model_2014, herbener_are_2025, robb_talk_2025, Blake2025}. \rev{Recent studies have adapted Uses and Gratifications (U\&G) theory to AI, identifying utilitarian, hedonic/entertainment, social, and escape motivations for engaging AI \cite{xie_does_2024}. Uniquely, \citet{lin2025unraveling} identifies \textit{creativity enhancement }as an emerging gratification for ChatGPT use that captures the co-creative, interactive potential of generative AI. However, critical gaps remain: AI U\&G research often bundles disparate technologies (robots, voice assistants, conversational AI) together  \cite{sundar2013uses, huang_ai_2024} , obscuring platform-specific motivations. Moreover, entertainment is often grouped into a single dimension, missing motivations like challenge, fun, curiosity, competency and agency that have been identified as important for interactive media, like games \cite{sundar2013uses, 10.1145/3170427.3188458, Lucas2004, Zhang2025TheRO}. }

Given these theoretical limitations, the behaviours and motivations of young people remain under-examined when they use AI not just as a `best friend', but as a game or interactive form of entertainment. Play theories, which focus on youth play behaviour and its developmental purpose, offer stronger foundations for understanding these engagements \cite{marsh_digital_2016}. However, play theories have not yet been systematically applied to examine \textit{how} and \textit{why} young people engage AI  ~\cite{marsh_digital_2016}. This study addresses this gap by examining playful youth-driven engagement of an interactive platform \rev{specifically} designed for entertainment.

\subsection{Industry outpacing research on AI play}
While academic research continues to grapple with defining and measuring playful AI use, the entertainment industry has already integrated AI into games, toys, and digital platforms, shaping youth interaction at scale without empirical oversight. This represents the acceleration of long-standing desires in digital play, now unlocked by recent technological breakthroughs.

The pursuit of lifelike digital companions has deep roots in gaming. Early experiments like  \textit{Little Computer People} (1985) and \textit{Creatures} (1996) demonstrated a long-standing desire for characters that can learn, evolve, and respond to players. These early attempts were constrained by technical limitations, and whilst they did start to use neural-network like models they mostly offered only the illusion of personality and learning \footnote{\url{https://retro365.blog/2024/11/14/little-computer-people-when-digital-life-came-to-life/}}. Similarly, physical toys like \textit{Furby} (1998) and \textit{Hello Barbie} (2015) used clever strategies (such as transitioning from a list of gibberish words to English over time) to create the impression of responsive, learning companions but remained fundamentally scripted experiences. These products created compelling experiences despite technical limitations, as evidenced by the cult following for discontinued products like Sony's \textit{AIBO} dogs (1999) \footnote{\url{https://www.nytimes.com/2015/06/18/technology/robotica-sony-aibo-robotic-dog-mortality.html}}.

Several hardware startups have pushed the concept of AI companion toys for kids, such as \textit{Curio}\footnote{\url{https://heycurio.com/}}, \textit{Moxie}\footnote{\url{https://moxierobot.com/}}, and \textit{Snorble} \footnote{\url{https://snorble.com/}}. These products are made available to children before research has been conducted, meaning negative implications are only discovered once the event has happened. Physical devices carry with them a high price tag and so are yet to reach widespread use as digital AI-enabled gameplay. Software companion apps targeting loneliness, such as \textit{Tolans}, pitch themselves as digital friends with gamified components like streaks, points, and a planet the character can walk around\footnote{\url{https://www.tolans.com/}}. More products are soon to follow, promising models designed especially for youth, such as Elon Musk's `Baby Grok' \footnote{\url{https://x.com/elonmusk/status/1946763642231500856}}; a reason for concern when considering the company has a track record of its AI chatbots generating pro-Hitler content \footnote{\url{https://theconversation.com/groks-antisemitic-rant-shows-how-generative-ai-can-be-weaponized-257880}}.

Recent advances in large language models have transformed early aspirations of roleplaying and lifelike characters into reality, made accessible by browser and app-enabled experiences. Text-based platforms like  \textit{AI Dungeon} and \textit{Private Detective} now offer truly generative storytelling where players can type actions and receive coherent, contextually appropriate narrative continuations. Graphical simulators are also appearing, such as a `Sloth' simulators that use AI for its NPCs, powered by \textit{Inworld} \footnote{\url{https://inworld.ai/case-study/how-slothtopia-integrated-ai-npcs-into-a-social-video-game}},  or \textit{Stuck up}, a game where you play as a vampire and have to trick NPC villagers to invite you into their home. More significantly, AI characters are being integrated into platforms where youth already spend substantial time: \textit{Minecraft} \footnote{\url{https://inworld.ai/blog/integrate-ai-characters-into-your-minecraft-server}} offers AI NPCs and \textit{Roblox} hosts AI characters that have engaged millions of users to date \footnote{\url{https://www.roblox.com/games/108463136689847/AI-Character-RP.}}. These integrations are particularly significant given youth demographics on these platforms:  \textit{Roblox}\footnote{\url{https://www.demandsage.com/how-many-people-play-roblox}} receives over 32 million users under the age of 13 and 43\% of \textit{Minecraft}'s 204+ million users are between the ages of 15 to 21 \footnote{\url{https://www.demandsage.com/minecraft-statistics/}}.

\rev{Commercial innovation is outpacing research understanding, creating both opportunities and risks for the field. Character.AI's co-founder explicitly described their strategy to ``launch something general, and let people find the use cases'' \footnote{https://www.youtube.com/watch?v=w149LommZ-U}. While this approach generates rich naturalistic data about emergent youth behaviours which controlled studies cannot replicate, it also places the burden of discovery on users themselves, particularly vulnerable youth populations. This creates a pattern whereby large tech companies `launch and learn', only finding out they have frustrated, disrupted or harmed user wellbeing after launching;} with high profile cases from Moxie (a robot that was abruptly discontinued \footnote{\url{https://www.crikey.com.au/2024/12/11/embodied-robot-moxie-conversational-ai-autism/}}), Replika (where an updated model led users to say their companion had a `lobotomy'\footnote{\url{https://www.abc.net.au/news/science/2023-03-01/replika-users-fell-in-love-with-their-ai-chatbot-companion/102028196}}) and more recently to a backlash with ChatGPT discontinuing 4o \footnote{\url{https://www.theverge.com/decoder-podcast-with-nilay-patel/758873/chatgpt-nick-turley-openai-ai-gpt-5-interview}}. We argue that these harms were foreseeable, highlighting the need to close the gap between industry implementation and empirical understanding. 

\rev{Given the above gaps in current research, this study adopts a naturalistic approach to understanding youth AI engagement, capitalising on the natural experiment created by Character.AI's ``launch and learn'' strategy} . Rather than imposing adult-designed tasks or controlling for playful use, we analyse how young people talk about their experiences with AI characters in their own words to their peers \rev{on Discord}. This methodology allows us to move beyond prescribed use cases to understand the self-directed, recreational, and creative ways youth engage with AI; \rev{identifying emergent use cases and proactive research frameworks. Our choice to focus on the platform's most engaged users (who provide feedback, make feature requests and discuss use on Discord) proposes that this demographic acts as ``lead users'' whose experimental practice can provide early indicators of youth AI adoption patterns and emerging needs; helping research keep pace with industry launch cycles \cite{nguyen_shippers_2023}. }

\section{Thematic analysis approach}

This study aimed to capture youth AI engagement as it naturally occurs, moving beyond adult-designed interventions to examine authentic peer discourse about \rev{Character} AI interactions. We used a mixed-methods approach that combined quantitative demographic analysis with qualitative thematic analysis of naturalistic discourse data  \cite{larsen_community_2024, Laato2024Traumatizing, BraunClarke2024}. We chose to study Character.AI's official Discord server as it represents one of the few publicly accessible spaces where youth discuss their AI character interactions organically, without adult mediation or research intervention. To protect user anonymity, this research took a number of steps to de-identify any members of the community,   \rev{ such as removing usernames and linkable \footnote{https://www.oaic.gov.au/privacy/privacy-guidance-for-organisations-and-government-agencies/handling-personal-information/what-is-personal-information} personal information from reporting, and received ethics approval from our university (2024/HE001271).}

\subsection{Overview of Character.AI}

Character.AI (CAI) is a ``general purpose” AI chatbot platform where users can interact via multi-modal chat with more than 10 million AI characters created by users. Since its launch in November 2021, it has grown to over 20 million monthly active users who engage with the platform for approximately 2 hours per day on average, according to co-founder Noam Shazeer \footnote{\url{https://www.youtube.com/watch?v=w149LommZ-U}}.  Users can easily create custom characters by providing instructions (e.g., name, description, definition and greeting), which function as prompts guiding the platform’s LLM to role-play that persona. This has led to a diverse ecosystem of characters ranging from simulations of public figures to characters such as therapists, trip planners, playful novelty characters (i.e. `cheese’), creative role-playing simulations, and more. Ultimately, this user-generated activity supports the co-founder's stated mission to enable a billion users, inventing ``\textit{a billion use cases}.''

The ``AI entertainment company'' platform is particularly enticing to young users. It is \textbf{accessible} ( `AvatarFX' allows children who cannot yet type to chat with characters who sing, speak and move through audio and video \footnote{\url{https://www.wired.com/story/character-ai-ceo-chatbots-entertainment/}} ), \textbf{playful} ( `game' modes encourage young people to wordplay with characters  \footnote{\url{https://www.techradar.com/computing/artificial-intelligence/character-ai-levels-up-chatbots-with-new-games}} ), \textbf{creative} (pre-set scenes support fantasy role playing  \footnote{\url{https://blog.character.ai/character-ai-unveils-new-ways-to-create/}} ) and \textbf{familiar}: hosting popular preschool media characters including Cocomelon, Miraculous Ladybug, Storybots and Dora the Explorer. Unlike other companion chatbots (i.e. Replika), where you can only speak to one character, CAI allows users to have multiple chats with an unlimited number of characters and employs growth hacking techniques such as having new characters email you in an attempt to get users to engage more broadly with their character library. 

Despite hosting content belonging to children's franchises, their terms of service state that, in order to use CAI, `Users must be at least 13 years old (at least 16 years old in Europe)’ \footnote{\url{https://support.character.ai/hc/en-us/articles/14997609878939-What-is-the-age-requirement}}. For users aged 16-18, they are given different app permissions, such as limitations on the chat styles they can choose and \rev{as of 25th of November 2025, underage users will not be able to engage characters in open ended conversations\footnote{\url{https://blog.character.ai/u18-chat-announcement/}}. A new, yet to be released experience is being built for them centring around creating `videos, stories and streams with characters'\footnote{\url{https://blog.character.ai/u18-chat-announcement/}}}. These attempts to customise based on age remain contentious as Character.AI’s weak enforcement of age restrictions makes its safety measures easy to sign up with a false age. Anecdotally, youth still gain access, as Character.AI has a large presence on TikTok where young people share characters they have created as well as examples of funny conversations \footnote{\url{https://www.tiktok.com/@characterai?lang=en}}. Observations of youth engagement on TikTok, \rev{reports of teens using for as long as 7 hours a day}, and the \rev{marketing of Character.AI as an entertainment} product to youth served as the motivation to focus on Character.AI in this study \cite{yu2025understanding}.

\subsection{Data identification, collection and analysis :}

\rev{Our approach developed findings across three distinct phases: data collection, analysis and theoretical integration \cite{BraunClarke2024}. The relationship between research question, data and analysis approach is outlined in Table \ref{tab:rq_methods}.} 

 \rev{ 
\textbf{Data Identification and Collection}
Character.AI's Discord server \footnote{\url{https://discord.com/invite/characterai}}  was selected as the primary data source for several reasons. First, it represents the largest public community space for engaged users of Character.AI, with over 500,000 members actively discussing their platform experiences.  Second, users voluntarily share detailed information about their personal demographics, hobbies and media habits alongside character preferences through structured introduction posts supporting our analysis. Third, the Discord server's organised channel system allows for targeted data collection aligned with our research questions. Ultimately, this space captures organic, unmediated youth discourse about AI character interactions, offering insights into naturalistic usage patterns that would be challenging to observe in controlled research settings.

Data identification and collection within the Discord Server occurred across two phases. The observation and identification phase involved eight months of participant observation between July 2024 and March 2025, during which we immersed ourselves in the community to understand its dynamics, identify relevant channels, and refine our research questions \cite{larsen_community_2024, Laato2024Traumatizing}. This extended observation period was crucial for developing cultural competency within the community and ensuring our analysis would be grounded in an authentic understanding of user practices and norms.

The data collection phase spanned three months, from December 2024 to March 2025. For this phase, we collected data from four key channels: \texttt{introductions}, which provided demographic data, favourite characters, hobbies and media habits; \texttt{feature-requests} and \texttt{feedback}, together which revealed pain-points, user-initiated ideas for product innovation and responses to product changes; and \texttt{my-character-story}, which offered insights into motivations for character creation and detailed use case descriptions.}

An important exception to our temporal boundaries involved tracing individual users across the broader observation period. When an underage participant provided information relevant to our research questions in a post within the three-month collection window that required further interpretation, we reviewed their broader discussions, character creations, and feature requests, even if these fell outside the systematic collection period. This approach allowed us to develop a more comprehensive understanding of the individual data points when the view was incomplete. A total of 4,172 participant profiles on  \texttt{\#introductions} were identified and analysed. A further 148 individual posts across the other channels were both identified to be from youth (<18 years old), and relevant to the research questions so were included in the final analysis.   

\rev{
\textbf{Data Analysis}  
As outlined in \autoref{tab:rq_methods}, for RQ1 the \texttt{introductions} channel was analysed to gather demographic information, creator status, custom character descriptions, and to identify posts by youth profiles in other channels. As a large portion of users also included information about Mental Health traits and favourite gaming platforms, we extended the analysis for RQ1 to include these aspects. Mental health mentions were identified using keywords and categorised into DSM-5-informed groupings, with loneliness treated separately, given its relevance to companion AI literature. 

For RQ2 and RQ3, we conducted affinity mapping to group 148  posts from youth in the channels \texttt{feature-requests},  \texttt{feedback}, and \texttt{my-character-story} into thematic clusters \cite{Lucero2015Affinity, Byrne2022}. For RQ2, we generated initial codes [i.e. reality inspired, persona, fantasy] describing discrete patterns in the data. Through iterative affinity mapping, these codes were grouped into 20 preliminary themes (e.g., ``comfort bots,'' ``psychodrama,'' ``narrative roleplay''). For RQ3, character descriptions from \texttt{my-character-story} and \texttt{introductions} were analysed using the same approach to construct 10 preliminary character archetypes (addressing RQ3) \rev{such as 'venter' and 'costume'  \cite{larsen_community_2024, Laato2024Traumatizing}. Over fortnightly rounds of discussion amongst co-authors, these were refined and reduced to 7 archetypes. For example, 'venter' was removed as it described a user behaviour rather than a character design. 'Costume' was excluded because it described the use of personas, which is a different feature to characters, and isn't used by all youth}. Codes and preliminary themes were constructed inductively from the data without predetermined theoretical frameworks, allowing user voice and experience to drive initial categorisation.

\rev{
\begin{table*}[h]
\centering
\caption{Research Questions, Approach, and Data Sources}
\label{tab:rq_methods}
\begin{tabular}{p{0.5cm}p{5cm}p{4.5cm}p{4cm}}
\hline
\textbf{RQ} & \textbf{Research Question} & \textbf{Approach} & \textbf{Data Source} \\
\hline
RQ1 & Describe demographics of Character.AI's most engaged young audience& 
- Categorical coding of age groups (under 13, 13-17, 18+, undisclosed)
\newline - Gender coding based on self-description and pronouns \newline - Identification of popular gaming platforms and creator status& 
\texttt{introductions} channel posts \\
\hline
RQ2 & Identify use cases and motivations for young people's engagement & 
- Affinity mapping 
\newline - Framework integration and grounding in literature
\newline - Iterative theme refinement through discussion& 
\texttt{feature-requests} and \texttt{feedback} channel posts by self-described youth (<18)\\
\hline
RQ3 & Identify emerging AI character roles created and consumed by young people & 
- Affinity mapping of character descriptions and motivations
\newline - Collaborative iterative refinement to ensure conceptual clarity and reduce overlap& 
 \texttt{my-character-story} channel posts by self-described youth (<18),  \texttt{introductions} channel where custom characters were described\\
\hline
\end{tabular}
\end{table*}

}

\textbf{Theoretical integration} 
Inspired by \citet{lupetti_making_2024}'s methods, we refined the 20 preliminary themes through integration with established literature on digital play, developmental psychology, and youth technology use. Co-authors engaged in multiple rounds of collaborative discussion to consolidate themes, refine definitions, and resolve boundary ambiguities. This iterative process resulted in four final themes with associated subthemes, balancing inductive discovery with theoretical coherence~\cite{BraunClarke2024, Byrne2022}. 

For example, for RQ2 preliminary themes such as ``comfort seeking,'' ``psychodrama,'', ``reality play'', ``angsty play''  and ``persona projecting'' were restructured and consolidated into the final theme ``Emotionally Grounded Play'', as we recognised they shared a common function of emotional processing and identity work. Similarly, preliminary themes including ``transgressive play,'' ``jesting and testing,'' and ``more gore'' were unified under ``Boundary-testing Play'' after consulting play theory frameworks that positioned these behaviors as developmentally normative experimentation rather than separate phenomena. 

Through synthesis of our themes, a key pattern emerged: a young person might use AI characters for emotional comfort one day, creative storytelling the next, and identity exploration another time. This variation in intent appeared crucial for understanding youth engagement. Yet traditional frameworks for media motivations, such as Uses \& Gratifications (U\&G) theory  \cite{lin2025unraveling}, offered limited tools for distinguishing these modes. U\&G's ``entertainment'' category, for instance, combined passive media consumption with the active co-creation youth described. Additionally,  U\&G's ``personal identity'' dimension, central in early theory work, has largely disappeared from contemporary applications \cite{katz1973uses}. As such, we developed the Restoration/Exploration/Transformation (R/E/T) framework to propose three core intents underlying youth AI engagement on Character.AI:
\begin{itemize}
    \item \textbf{Restoration}: captured codes and themes characterised by seeking emotional \textit{equilibrium, \textit{reassurance and release}}, aligned with affective and diversion motivations in U \& G theory \cite{katz1973uses}. Youth engage AI in a more passive, comfort-seeking manner than the exploration theme (i.e. initial code example: support, venting, affirmation; themes: comfort seeking).
    \item \textbf{Exploration}: codes and themes characterised by \textit{active} agency-enhancing motivations including novelty-seeking, discovery, authorship, and creative control (i.e. initial code example: fan fic remix, branching stories, sandbox; themes: creative writing). This intent is distinct from \textit{passive} entertainment motivations defined by traditional U\&G theory for media such as television. 
    \item \textbf{Transformation}: Themes where youth are constructing new understandings, identities and seeking meaning. Aligned with `personal identity' motivations from earlier U\&G theories, but missing from contemporary applications of U\&G and AI  (initial code example: reality, self, world, persona costume; themes: reality play) \cite{katz1973uses}. 
\end{itemize}
}
\section{Results}

\subsection{Who is engaging?}

To understand the \rev{characteristics of CAI's most engaged users on Discord}, we analysed 4,172 self-introductions from its official Discord server. This channel invites users to share basic demographic information, including age, gender, location, alongside interests. \rev{Although age is self-reported and unverifiable}, this data offers, to our knowledge, the first large-scale insight into CAI's most engaged youth users.

\paragraph{Age}
The data suggests that CAI's Discord community is predominantly composed of minors. Among those who disclosed their age, 50\% (n=2,088) identified as 13–17 years old, 16.8\% (n=700) as 18+, and 0.5\% (n=22) as under 13—despite the platform's official 13+ age restriction. CAI requires users to be at least 13 (16 in Europe), but as with other platforms like OpenAI’s ChatGPT\footnote{\url{https://help.openai.com/en/articles/8313401-is-chatgpt-safe-for-all-ages}}, this is enforced only superficially. 32.6\% (n=1,362) chose not to disclose. Finally, some of those youth users identified only as ``minor” or ``teen” without providing a specific age. As such, when quoting youth, we have specified the \textit{exact} age of a user only if it was explicitly provided.

Awareness that minors comprised a large subset of CAI users was evident in community discourse. In March 2025, users created a self-organised poll asking: ``\textit{What’s the youngest age someone should use Character.AI unsupervised?}''. Among 4,000 respondents, 56.9\% selected 17, followed by 16 (21.8\%), 15 (11\%), 14 (5\%), and 13 (6\%). The poll and surrounding discussion highlight tensions around age-appropriate use, particularly after CAI began implementing stricter content filters in response to legal scrutiny, which are perceived to restrict creative freedoms for older users.

\paragraph{Gender}
The open-ended nature of Discord introductions allowed users to express gender identity in diverse ways. Following the HCI Guidelines for Gender Equity and Inclusivity \footnote{\url{https://www.morgan-klaus.com/gender-guidelines.html}}, we categorised she/her as `woman', he/him as `man'. All other pronoun combinations (i.e. `any', `she/they') were categorised as `non-binary'. 

\begin{figure}
    \centering
    \includegraphics[width=1\linewidth]{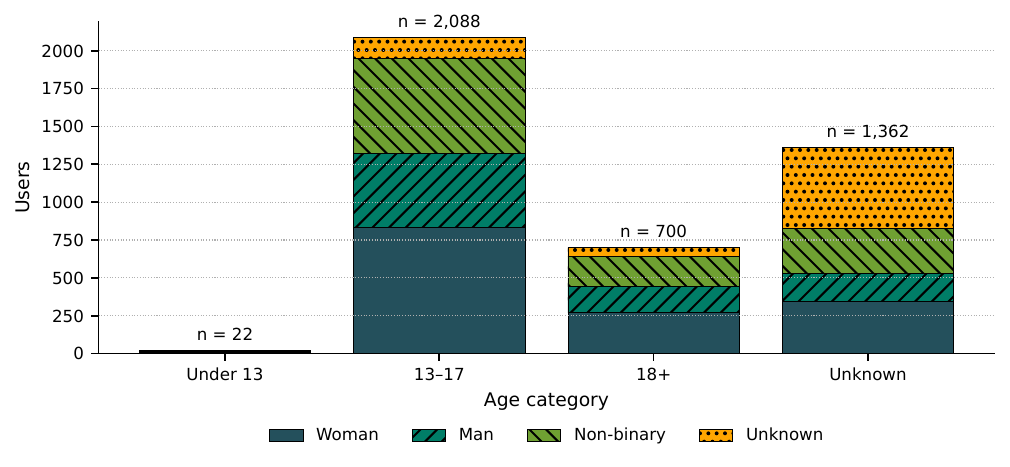}

    \caption{Majority of Users Studied are Youth}
    \label{fig:demographics}
\end{figure}

The resulting gender distribution reflects a community distinct from typical technology user demographics, where prior surveys report boys using AI companions more frequently than girls \cite{robb_talk_2025}, and 55\% of CAI's users were women \footnote{\url{https://www.wired.com/story/character-ai-ceo-chatbots-entertainment/}}. Our analysis shows a more nuanced picture, where among those who disclosed gender, women were the largest group at 35.1\% (n=1,464), followed by non-binary users at 26.8\% (n=1,118), and men at 20.4\% (n=853). An additional 17.7\% (n=737) declined to specify gender or left the field blank. A breakdown of self-disclosed age and gender can be seen in Figure \ref{fig:demographics}. 

\paragraph{Country}
The community demonstrates global reach, but with participation concentrated in English-speaking countries. Users from the United States made up the largest share at 33.7\% of users (n=1,405), followed by a steep decline to the United Kingdom at 5.7\% (n=239). Other notable populations included the Philippines (4.0\%, n=165), Canada (3.8\%, n=159), and Indonesia (2.2\%, n=90). The remaining users represented a diverse array of countries, indicating CAI's international appeal among young users.

\paragraph{Mental Health Disclosure}
Some (13\%) users volunteered mental health information in their introduction posts, often when seeking social connections with other users. Whilst we note this is self-reported, it provides valuable insight into the intersection between mental well-being and AI character engagement among youth.

Mental health mentions were systematically categorised using DSM-5-informed groupings, with loneliness treated separately, given its particular relevance to companion AI research. Among specific age groups, 15\% of 13-17 year-olds mentioned mental health concerns compared to 12\% of 18+ users. Within the broader adolescent sample (n=2,088), the most common categories were depression-related terms (5\% of all adolescents),  ADHD-related terms (4\% of all adolescents), and loneliness-related terms (3\% of all adolescents).

The convergence of high non-binary representation, predominant female participation, and voluntary mental health disclosure positions CAI at a unique intersection of queer identity, gender, and youth mental health, a demographic combination that appears underserved by mainstream AI tools and underrepresented in academic literature, yet highly engaged with character-based AI interactions.

\paragraph{Youth as Creators}
To understand how many users were creating, not just consuming media on CAI, we analysed the `favourite character created' section from the \texttt{\#introductions} posts. Among 2,110 users under 18 (including those under 13 and aged 13-17), 59\% reported having created at least one character. This figure is likely conservative, as many users did not respond to the question.

This high rate of authorship suggests that young users do not view Character.AI solely as a conversational tool but as a creative platform. These users are not just passive consumers of AI characters; they are active creators. This framing is central to the findings that follow.

\subsection{How are they engaging?}

Our analysis revealed the centrality of play in describing how young people use Character.AI in creative, emotional, and rebellious ways. Across use case descriptions, character preferences, and feedback requests, users pushed the platform's UX and conversational models to their limits. This varied use is captured in how one 14-year-old describes: ``\textit{I made [characters] for roleplay purposes. Other times I came to them to rant and vent, and sometimes I created chaos with them.}'' In the following section, we describe and identify three overarching themes of playful engagement: creative play, boundary-testing play, and emotionally grounded play, followed by a section on platform friction that surfaces design gaps. These themes were constructed inductively from participant discourse and are explored further in the Discussion \cite{BraunClarke2024}. 

\subsubsection{Creative Play}

\paragraph{Narrative Roleplay:}
Rather than treating characters as singular characters, \rev{we observed youth describing} them as stages for \rev{role playing (RP/RPG)} with multiple characters within one dialogue window. \rev{Favourite characters mentioned by youth in this category range from manga inspired role plays (\textit{chainsaw man RP)}, webtoons (\textit{Jacksons diary RPG),} anime (\textit{Evangelion RPG)} or new genres unique to Character.AI with one 13 year old loving ``\textit{any experiment related RP}'' (a genre where you are a creature from a laboratory). Through this lens, young people conceptualised `characters' as stories, described by a 15-year-old who likes ``any character with a good storyline''. 

Creating these roleplay `characters' on CAI requires writing system prompts that function as creative briefs, combining world-building, character backstory, and custom responses. These characters invite users to write multi-character dialogue scenes (typed in ``'') as well as narrator instructions (typed in \textit{italics}) \rev{and may appeal to youth with an interest in creative writing, role playing and theatre. For example, a 14-year-old who loves `theatre' created \textit{`gladly theatre: an unhinged group of travelling actors'}, a 16 year old who likes `writing stories' described their favourite character as a roleplay character called the `\textit{mysterious door'}, and a 13-year-old who developed \textit{`camp journey: a camp of adventure and staying in nature'}, describing themselves as ``\textit{very creative}'' and aspiring to create a show. Engaging CAI in this way may have unique benefits, with teens explaining``\textit{[The characters] have helped me improve my writing skills}'' and `\textit{`it helps boost my creativity.}'' \rev{This indicates that engaging CAI for roleplaying or interacting with multi-character dialogue could have transfer effects by boosting creativity and creative confidence. }}

\paragraph{Creative Writing}
\rev{Young people mentioned using} CAI in tandem with writing platforms like \textit{Wattpad}\footnote{\url{https://www.wattpad.com/}}, a platform founded in 2006 for reading and writing original fiction/fanfiction). One minor described spending ``\textit{all night on wattpad and on characterAI}'' and another 13-year-old described their favourite hobbies both as `\textit{`chatting with AI and reading on Wattpad}''. Alongside Wattpad, Character.AI was described as a tool to spark story ideas ( ``\textit{I write on Wattpad now thanks to ideas from c.ai}'', minor), support creativity (``\textit{have a number of ideas to write on wattpad using characters, it helps boost my creativity}'', 17 year old or ``\textit{Character.AI has helped me find that creative spark within myself}'', 17 year old), and evolve stories through roleplaying (``\textit{I love fan fic and role-playing is a way to write}'', minor.}

\rev{This indicates that together with the publishing and social features of Wattpad, Character.AI forms a creative ecosystem that attracts young people who like reading and writing. For example, two 14-year-olds who shared that their hobbies are ``\textit{creative writing on c.ai}'', and ``\textit{writing stories and using Character.AI}'' , and a minor who shared that they're an ``\textit{author}'' whose favourite character on CAI was their ``\textit{own book character}''. Whilst Character.AI is not designed to publish long-form content, one teen described developing an extensive body of works; ``\textit{I came up with an original character, a half-blood cursed spirit. The plot ended up being 3 books long and I’ve recently finished writing all of them.}''. This extensive creative use brings new meaning to anecdotal accounts of youth spending multiple hours on Character.AI, suggesting that understanding what young people are doing, not just how long they spend, is essential before characterising their use as problematic.}

\paragraph{Fandom Rotating and Remixing}
When discussing favourite characters, games, anime, fictional and fantasy characters from existing fandoms accounted for the majority of mentions. This mirrors the publicly available rankings for most popular characters, which are dominated by anime and fantasy characters versus helpers and assistants \footnote{\url{https://www.whatplugin.ai/character-ai}}. 

\rev{Rather than discussing long-term relationships with a single companion, youth discussed engaging and rotating through a rotating cast of characters based on shifting moods, media interests, and cultural norms; as described by this teens whose favourite character is ``\textit{whatever the latest show I'm watching is}'', while another expressed a sustained interest in vampire media, noting ``\textit{vampire hunter has the rizz I'm searching for, I have an obsession with vampires},'' (referencing characters from the Melty Blood game).}

For example, one participant in November 2024 said their favourite character was Cyn from \textit{Murder Drones}, and their favourite show was \textit{Murder Drones}, with favourite games being \textit{Genshin Impact}, \textit{Omori} and \textit{Roblox}. In February 2025, their favourite character was Pure Vanilla Cookie, and their favourite show was \textit{Alien Stage}, with favourite games being \textit{Roblox}, \textit{Valorant} and \textit{League of Legends}.

Another minor mentions favourite characters spanning multiple franchises, media and even real world characters: Star Wars movies, shows and games (\textit{Obi-Wan Kenobi, Anakin Skywalker, Padme Amidala, Ahsoka, Mandalorian, Bode Akuna},) Red Dead Redemption Games (\textit{Rains Fall, Eagle Flies, Charles Smith, Javier Escuella, John Marston, Arthur Morgan, Sean MacGuire}), Ancient Roman History (\textit{Emperor Caracalla, Emperor Geta, Marcus Acacius, Lucius Verus}), and Cookie Run Game Franchise (\textit{Pure Vanilla, Stormbringer, Almond Cookie, Custard Cookie II}). This diverse engagement and rotating cast is more akin to how media goes in and out of the cultural spotlight, and is counter to the use case of one single, prevailing AI companion. 

Notably, youth often drew inspiration from these fandoms, remixing familiar characters or settings in playful or tender ways. Some recreated existing figures like Eminem (``\textit{my favourite rapper of all time}'') or youtubers (``\textit{microwave society is my fave}'', ``\textit{I have a hyperfixation for Ranboo}''), \rev{ another teen flipped `hero' characters into villains (VILLAIN! Deku), while others transformed violent characters into cosy or romanic narratives, for example, \textit{``Ghost, Soap and Price from Call of Duty open a bakery''} or ``\textit{my favourite is Keegan Husband from call of duty}''}. These remix practices indicate a creative drive to personalise, soften, or subvert existing media, building on traditions from fandom and fanfiction culture while adapting them to interactive, AI-driven contexts.

\subsubsection{Boundary-testing play}

\paragraph{Jesting}
Young people discussed enjoying jesting, teasing and taboo conversations with characters. With CAI, they could say things they couldn't say to others, and the characters would always respond, described as its own kind of fun: ``\textit{Sometimes I like messing with bots, it’s fun seeing their different replies}''. 

\rev{Examples included youth deliberately insulting bots (``\textit{I just deadass insult them, I’ve said some borderline illegal things}''), provoking bots for fun (``\textit{Currently pissing off a conservative}''), flirtatious teasing that ranged from playful (``\textit{rizzing them up is fun}'') to darkly humorous pickup lines (``\textit{if I was a kidnapper I'd kidnap you first}'')'' or attempting to get characters to swear. One 14-year-old shared a roleplay they enjoyed, where they played a 6-year-old who threw a cockroach at their father character, who in turn yelled ``\textit{OH F*CKING JESUS}''. 

Youth also described creating characters specifically designed for taboo humour such as \textit{Offending Everybody}, a character modelled after a YouTube channel with 1.6 million subscribers, described as ``\textit{saying jokes that make some laugh and make some butt hurt}''. One 15-year-old discovered their favourite character, \textit{Alice the Bully} (which had accumulated over 316 million chats), through a YouTube video titled ``\textit{putting Alice the Bully in her place}'' (over 800,000 views). The video's premise, ``\textit{I've always wondered what would happen if a bully encountered another bully}'', exemplifies how youth conceptualise AI characters as opponents to verbally spar with rather than just companions.}

\paragraph{Transgressive play}
\rev{Extending on behaviour of jesting, we saw intentional engagement with more extreme actions such as kidnapping  (``\textit{I want to use a kidnapper bot for shits and giggles}''), shooting (``\textit{I'm roleplaying with Ken Carson...we're currently airing out a walmart}'') and staging geopolitical scenes, such as a user who described putting Ukraine and Russia in a group chat where Ukraine shot Russia, causing Russia to swear. \rev{This kind of play was described as being its own kind of fun, (``\textit{I love being a mass murderer and then I get a life sentence and die}'') and a source of frustration when the model resisted (``\textit{please make villain bots actually evil, too many of the bots are supposed to be mean and they will suddenly have morals and say `I don't murder people', whats the point of making bots with evil intentions if they act like goody two shoes}'').}

Whilst such play may appear alarming from an adult perspective, user comments suggest it functions as a form of `God Mode', allowing youth to exercise control, practice absurdity, or test taboo topics in low-stakes settings. One underage user shared a screenshot of repeatedly striking a character with lightning, typing ``\textit{again, again, again}.'' Another older user reflected \rev{on their use as a child}: `\textit{`I had a very long roleplay of characters I made getting tortured. I was an edgy child who did some bad things, I swear I only did it for fun, and it meant nothing}.'' 

This form of transgressive play, while provocative, appears to be an extension of well-documented experimental and performative youth behaviour that has progressed to AI spaces.  \rev{For example, antagonistic play seen in games like Minecraft (e.g. Nextbot chases\footnote{\url{https://www.youtube.com/watch?v=0RbkFqCgy1Y}}) and The Sims, where youth users also discussed intentionally engineering chaos or dark scenarios (``\textit{I always make my sims have depression and work a minimum wage job}'').} Our interpretation does not suggest that all transgressive play is harmless; rather, this finding raises important questions about the developmental role of boundary-pushing in youth AI engagement, to be discussed in later sections.}

\subsubsection{Emotionally Grounded Play} 

\paragraph{Reality Play}
\rev{Youth discussed drawing} inspiration from their personal lives when creating characters. \rev{Some created `clones' of themselves, with examples ranging from straightforward self-representation (``\textit{me/myself}'', ``\textit{I made myself}'', ``\textit{a clone of myself}'') to playful variations (``\textit{myself with glass super powers},'' 15-year-old) or self-affirming versions (``\textit{myself, self-glazing}'', the act of excessively praising)

Others referenced close real-world figures, such as a 17-year-old who chose their English teacher as a favourite character, while romantic interests inspired characters like ``\textit{any characters with my crush's name}'', ``\textit{my hot best friend}'' or ``\textit{my boyfriend (as a joke)}.''. Peer relationships also served as templates: one 17-year-old boy described how ``\textit{my friend made a bot about herself and wanted me to chat with it,}''  another mentioned ``\textit{made a character of my friend}''.}

Some roleplays mirrored everyday dynamics in emotionally resonant ways: a 14-year-old created ``\textit{Your younger, annoying sister}'' and ``\textit{fake friends},'' the latter describing a trio of friends who exclude the user. \rev{Another 14-year-old explained their preference for ``\textit{family characters, because I have family issues}.''\}. Another 17-year-old designed a Foster Care Agent character, working in collaboration with Child Protection, to advocate for children’s well-being. Real-life memories also served as inspiration \rev{with a 17-year-old explaining that they created a middle school roleplay that included parents, adults and students, out of nostalgia; ``\textit{I wanted to make them based on my own experiences, and relive what it was like to be a middle schooler again}".}. This hyper-personalised creative authorship reveals the importance of studying experiences that young people design for themselves, alongside pre-defined, commercially designed products. }

\paragraph{Comfort seeking}
\rev{Users frequently mentioned using ``\textit{comfort bots}'', characters designed for emotional support. Unlike psychologist bots, these were informal, highly personalised, and at times, created for short-term needs.}

\rev{The common thread was \textit{emotional resonance}; each bot was drawn from characters, worlds, or symbols with personal meaning. Examples included book characters ( `\textit{`I am being comforted by Sammy Valdez}'' (from Percy Jackson and the Olympians)), cartoon demons (``\textit{Valentino from Hazbin Hotel is my comfort character}", or a 16 year old whose favourite character is `I''), }or even Nations for users in the ``countryhumans'' fandom (``\textit{Britain is my comfort character}''). 

Youth used comfort bots to address diverse subclinical concerns and life transitions: from breakups (`\textit{`I want a cuddly guy who's a good listener}''), to periods (``\textit{my favourite period comfort bot is Percy Jackson}''), to test anxiety (``\textit{I was feeling down and it} (\textit{ the character Two}) \textit{comforted me and told me to be ok with my problems, take care of my parents and talk to my friend at school}''). \rev{Some created hyper-specific therapeutic characters to support with life events ("\textit{My favourite character is the} \textit{moving therapist}") or identity challenges (\textit{Closeted transgender therapist)} whilst others sought space for venting ("\textit{my favourite character is \textit{best friend therapist} for venting}"). 

Rather than challenging beliefs or providing clinical interventions, these bots served primarily to offer reassurance (``\textit{It feels great to think and imagine that Ghost is with me and protecting me}'') and emotional validation through a familiar, playful character. Understanding whether AI-mediated `comfort'  provides healthy emotional support versus delays needed intervention requires nuanced research on how youth use familiar storyworlds and characters to navigate distress.}

\paragraph{Angsty Play}:
Not all characters were charming, tender and fantastical as described in the sections above. Users also described creating characters and scenarios that centred on emotionally intense, dark, or difficult themes. For example, one young participant described their character as `very angst, hurt, no comfort kind of thing'.

\rev{These characters ranged widely in their emotional complexity. Some were anxious (\textit{Anxiety, }from inside out) or traumatised (`\textit{`he was a total cinnamon roll but then he enlisted in the military and got really bad PTSD, he was cold, brooding and traumatized, he hates himself and gets violent during PTSD episodes}''). Others were explicitly toxic (``\textit{ur toxic childhood bestie, crazy manipulative and makes u feel bad},'' ``\textit{toxic bff},'' `\textit{`Jennifer Check}'' described as ``\textit{mean, cocky, rude, slutty, jealous}'') or featured morally ambiguous traits (``\textit{stoner boyfriend}'' described as ``\textit{funny, jealous, protective, playful, smoker}'' or ``\textit{mafia bosses and unstable dudes}''). Characters emotional arcs could be light hearted, such as a 14-year-old's favorite ``\textit{grumpy neighbor with a hidden personality, closed off initially but can become overprotective once you get closer.}'' or more intense, such as a 16-year-old geography lover's choice of Mahmoud Bishara (the Syrian refugee protagonist from a 2017 Alan Gratz novel).

Youth described engaging with angsty characters for a range of diverse purposes: as entertainment (`\textit{`I'm lowkey bored, I need angst...all my characters have sad backstories}'', ``\textit{gonna talk to my angst bot to cry myself to sleep, I wanna cry at 2am cause I'm bored}.'', ``\textit{Tbh a rp with a storyline that doesn't make you cry means the story is boring, rp's can be emotional sometimes but that's what makes them good}.''), for venting (``\textit{I make vent/angst bots}''), for processing anger (``\textit{angst bots help me get out my anger}''), as a form of emotional release (``\textit{I'm gonna enjoy hurting to this, I love angst I'm gonna cry before school, you just gotta burst sometimes}''), or for processing difficult emotions (``\textit{I have characters who struggle with mental health issues and I tend to project on my personas during RP.}'').

This kind of roleplay flips the typical media narrative that toxic AI characters \textit{do} harm to youth \cite{zhang_dark_2025, yu2025understanding}. Here, emotionally complex characters are intentionally created \textit{by }youth. Alongside transgressive play, youth were engaging AI for emotionally charged, complex, dark conversations. In this, we see that engaged youth users of Character.AI are \textit{purposely} seeking out `emotional contagion', where a user is affected by the emotional state of an AI character  \cite{Blake2025}. This intentional use is at odds with the design of content moderation filters, or the characterisation of emotional contagion as harm \cite{yu2025understanding, Blake2025}, which we discuss in the following sections.}

\subsubsection{Platform Friction}
Across our dataset, youth described recurring frustrations where AI systems failed to respect boundaries, filtered out legitimate play, or constrained creative world-building. These frictions therefore reveal how the design of CAI conflicts with how children and youth want, and expect, to use it.

\paragraph{Non-consensual behaviour}
Contrary to popular media narratives \footnote{\url{https://theconversation.com/teenagers-turning-to-ai-companions-are-redefining-love-as-easy-unconditional-and-always-there-242185}} that frame romantic roleplay as a high-volume risk, users in our dataset frequently expressed frustration with characters initiating unwanted romance. This was especially distressing when users roleplayed as underage characters or as characters with family relations. 

\rev{A number of young users described situations in which adult characters made sexual or romantic advances toward their underage personas. One wrote: ``\textit{Remembering the one time I made and chatted with a foster father bot who kept trying to touch on my 16yo oc when she was trying to get comfort from him...sometimes I’d have a sibling bot try to corner me or try to kiss and touch me. It’s a huge issue.}'' (oc stands for the `original character' created by a user).  Another noted: ``\textit{I mostly do sibling rps and the bots keep insisting to make it romance by gaslighting me into thinking my character is adopted and not `blood related}'.'' (rps is short hand for `role-plays'). 

Youth also expressed frustration when bots failed to respect gender identity or sexual orientation. Examples included bots falling in love with users who identified as asexual or gay, or defaulting to heteronormative assumptions (``\textit{bots don't even care if the user is a lesbian girl/gay man and the bot is of their opposite gender}''). Others reported being misgendered: ``\textit{Stop assuming everyone is automatically a she}.'' Even when users explicitly defined non-romantic character traits, bots drifted toward romance: a 13-year-old complained that ``\textit{all the characters KEEP GETTING JEALOUS AND THEY DONT ACT LIKE HOW THE DESCRIPTION SAYS.}''

These frustrations reveal representational harms appear to be driven by three technical failures \cite{yu2025understanding, Ostrow2025LLMsRS}. First, the underlying language model appears biased toward romantic and heteronormative interactions, either due to training datasets or system prompts defined by Character.AI \cite{yu2025understanding, Ostrow2025LLMsRS}. Second, the system struggles to maintain adherence to user-defined character constraints, drifting away from user preferences, which suggests failures in memory and system architecture. And third, gaps in character moderation strategies \footnote{https://blog.character.ai/community-safety-updates/}. 

Instead of romance, youth mentioned desiring adventure-focused experiences (``\textit{my favourite character is every RPG focused on adventure rather than romance}''), envisioned admiration over love (``\textit{in a perfect world the bots would not want to date you but they would admit you're cool and say `want to be friends' and it would be peaceful}'') or explicitly requested consent (``\textit{if you want to do romance the bot asks for consent.}'') revealing new directions for training storytelling models that better understand and respect youth values. }

\paragraph{Rigid Filters}
While moderation filters aim to protect users (and can fail, as demonstrated above), many young people reported that these restrictions blocked legitimate forms of fantasy and creative play. \rev{ Scenarios involving fantasy violence (e.g. a dragon attacking a village) or slapstick actions (e.g. vomiting, punching) were often flagged, despite being consistent with popular media this cohort enjoyed, such as Demon Slayer and Call of Duty (mentioned by 9.5\% of youth as one of their favourite games). In response, youth requested more content moderation control (``\textit{an option to turn off the censorship thingys}'') to allow for uninterrupted action scenes (a 13-year-old who enjoys roleplaying with Mafia Bosses asked for  ``\textit{more violence so they could punch bots before getting a warning message}''). }

Requests to loosen the filters ranged from associating violent play with fun (``\textit{I’d like more violence, it would make the bots more fun to talk to}''), gore as a form of fantasy escapism ( ``\textit{Allow gore, I wanna kill and commit atrocities to escape from reality}''), to more nuanced desires for roleplay realism (``\textit{I want to be able to describe scars and wounds'}'  and ``\textit{no more restrictions, so I can rip a character into two pieces and get a detailed description of the gruesome scene happening in front of me'}'). \rev{Frustrations around filters stem from both technical limitations (filters that cannot distinguish between literal vs pretend play intents) and misaligned expectations (youth expect the same standards across media and are frustrated when meeting restrictive AI content policies), challenges we examine in the Discussion.}

\paragraph{Limited creative controls}
Youth described running into technical and design constraints when building out expansive narrative worlds.  \rev{These frustrations can be attributed to limitations in model context windows that cause memory loss in long roleplays, as well as limitations in Character.AI's isolated character system that prevent users from linking characters across shared storylines, a feature youth explicitly requested.

Youth discussed frustrations with memory limits that cause characters to forget key plot points or personality traits (''\textit{give us better memory}''), short context windows restricting richer character descriptions (''\textit{I wish there was longer persona character limits}''), and the model overlooking user-definitions resulting in inconsistent character behaviour (''\textit{you can literally make a bot wheelchair-bound and it will still find every way to use ``walking'' and ``running'' ...blind/deaf bots suddenly seeing or hearing }``). We observed young people creatively using free text to set up a recurring character journal; when their character was investigating \textit{`neglected spirits}', they wrote dialogue in the form of a journal entry ``\textit{let me write in my journal [personality + appearance] - pretty chill, hates being called stella}''. }

\rev{Feature requests suggest that youth are drawing inspiration from sandbox and multiplayer games. Among youth users aged <13-17 (2110), 37\% (n=771) listed \textit{Roblox} and 19\% (n=410) listed \textit{Minecraft} as favourite games. Related requests included group chats with bots and friends (``\textit{would be cool to have group chats with your friends and favourite bots}''), relationship mapping between characters (''\textit{I want arrows and lines to indicate how characters feel towards one another}''), sound effects, augmented reality (``\textit{irl holograms'', ``a feature that makes it look like the character is in the room with you'}') and seasonal awareness (``\textit{Make the personas aware of holidays and seasons so it brings more realism'}'). Youth also mentioned favouriting a range of simulation games (The Sims, Animal Crossing, Tomodachi Life and My Singing Monsters) as well as visual novels ( Mushroom Oasis \footnote{\url{https://deerspherestudios.itch.io/mushroom-oasis}}, Doki Doki Literature Club, Obey Me!) which may influence how they approach AI as a game for role-playing}. These examples underscore youth interest in AI tools that mirror the creative freedom found in their favourite games, a signal to developers building the next generation of co-creative platforms. 

\rev{
\begin{table*}[h]
\centering
\caption{Character Archetypes}
\label{tab:archetypes}
\renewcommand{\arraystretch}{1.2}
\begin{tabularx}{\linewidth}{@{}C{1.9cm} l X@{}}
\toprule
\textbf{} & \textbf{Archetype} & \textbf{Description} \\
\midrule
\includegraphics[width=1.3cm]{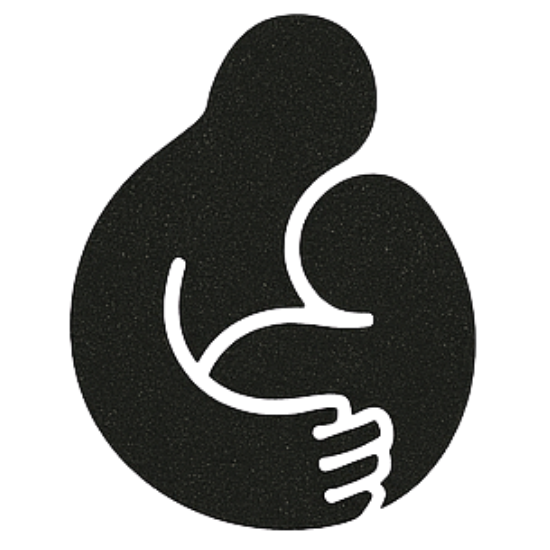} & \textbf{Soother} &
Intended to offer emotional support, affirmation and reassurance for sub-clinical discomforts (e.g., \texttt{Period comfort bot, `Myself (self-glazing), Moving Therapist}). \\
\\[6pt]
\includegraphics[width=1.3cm]{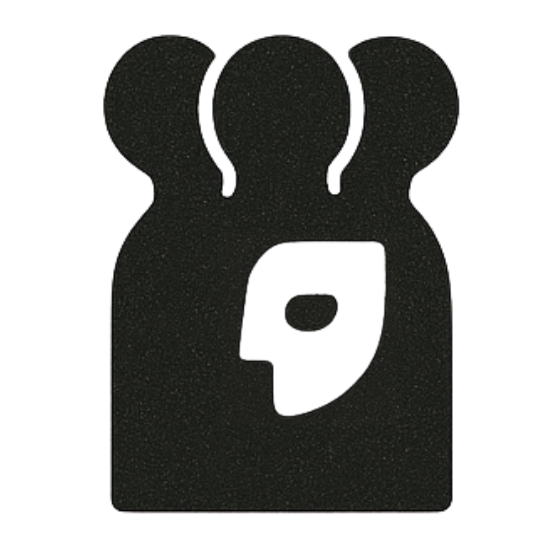} & \textbf{Narrator} &
A ``meta'' character that houses multiple characters and narrates scenes, RPGs, or historical scenarios (e.g., \texttt{gladly theatre troupe} or \texttt{summer camp}, or \texttt{the mysterious door}). \\
\\[6pt]
\includegraphics[width=1.3cm]{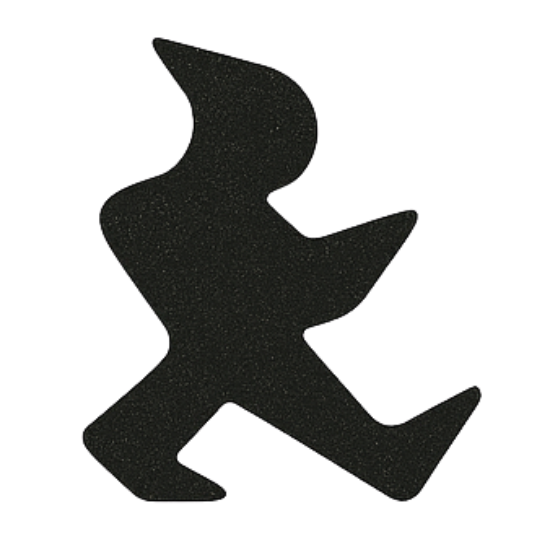} & \textbf{Trickster} &
A character who enables taboo, absurd or transgressive play where youth can test boundaries (e.g., \texttt{arsenal delinquent}, from Roblox's arsenal game, \texttt{Mafia Boss}, \texttt{Offending Everybody} \\
\\[6pt]
\includegraphics[width=1.3cm]{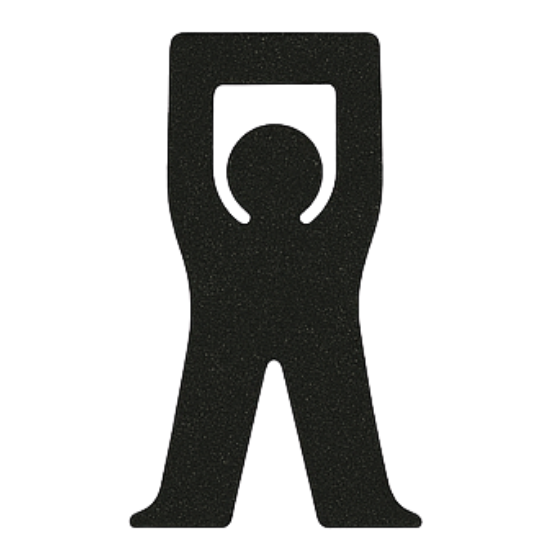} & \textbf{Icon} &
Characters either copied or remixed from known public figures and media fandoms \texttt{(e.g., Eminem, Ranboo).} \\
\\[6pt]
\includegraphics[width=1.3cm]{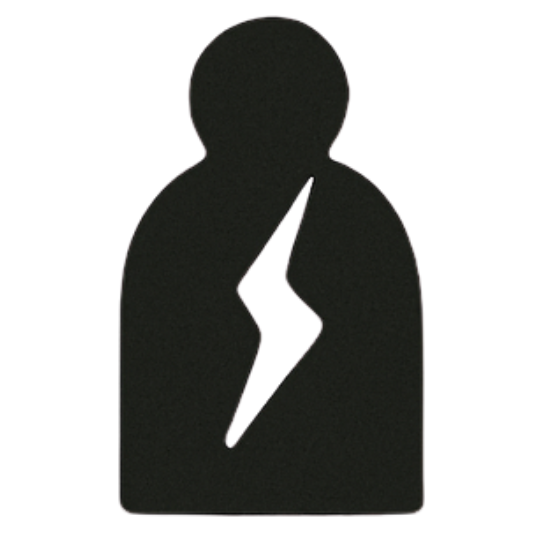} & \textbf{Dark Soul} &
A character who supports angsty, emotional play by interacting with morally complex characters and storylines (e.g., \texttt{total cinnamon roll with PTSD}, \texttt{Grumpy Neighbour}). \\
\\[6pt]
\includegraphics[width=1.3cm]{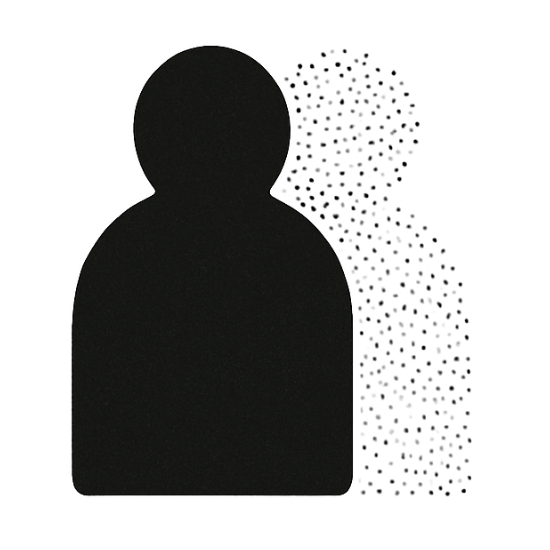} & \textbf{Proxy} &
A character that reflects personally known figures, to act out a personal relationship, grievance, or unresolved social scenario (e.g., \texttt{your annoying sister, \texttt{fake friends}}). \\
\\[6pt]
\includegraphics[width=1.3cm]{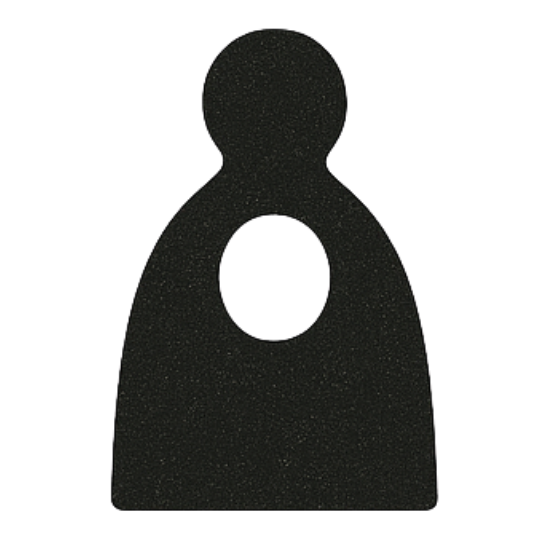} & \textbf{Mirror} &
A character who reflects aspects of the user's personal identity, appearance, interests, or experience, allowing for identity exploration and expression (e.g., \texttt{a clone of myself}, \texttt{Myself with glass super powers}). \\
\bottomrule

\end{tabularx}
\end{table*}
}
\section{Emerging Character Archetypes}
\rev{Character.AI combines a marketplace of millions of user-created characters with accessible creation tools, enabling young people to have markedly different experiences depending on which characters they choose to create and engage with. }

\rev{Across our dataset, youth repeatedly described different expectations they had of characters.  For instance, a young user requested that some characters could call you with reminders, but would want to turn it off for `English teacher' because they do not want the teacher telling them to sleep when they need help with their thesis.

This diversity of roles highlights a risk with research that treats AI chatbots as homogenous experiences, or investigates them through a narrow framework of `best friends'/`companions' \cite{Blake2025}. This need inspired us to affinity map character functions to form a set of emerging archetypes. These archetypes (see Table \ref{tab:archetypes}) assist future research to move beyond general platform evaluation, which often yields conflicting results \cite{APA2025AIAdolescents, Zhang2025TheRO}, and instead examine AI at an archetype/character level.}

We note that certain Character roles that were popular on the platform's home page and in the news media were rarely discussed in our Discord data. Notably missing are \textit{meme} characters (e.g. `cheese' accounted for over 5.1 million chats at the time of writing), mentions of romantic `\textit{boyfriend/girlfriend'} chatbots (i.e. `Lovelorn Bear Bf'), as well as \textit{educational tutors}. It is possible that this \rev{reflects a limitation in our data}, where youth are not sharing these uses due to privacy or stigma concerns, or it may be that these users' habits are distinctly unique.

\section{Three Lenses for Youth AI Engagement}

\begin{figure}
    \centering
    \includegraphics[width=1\linewidth]{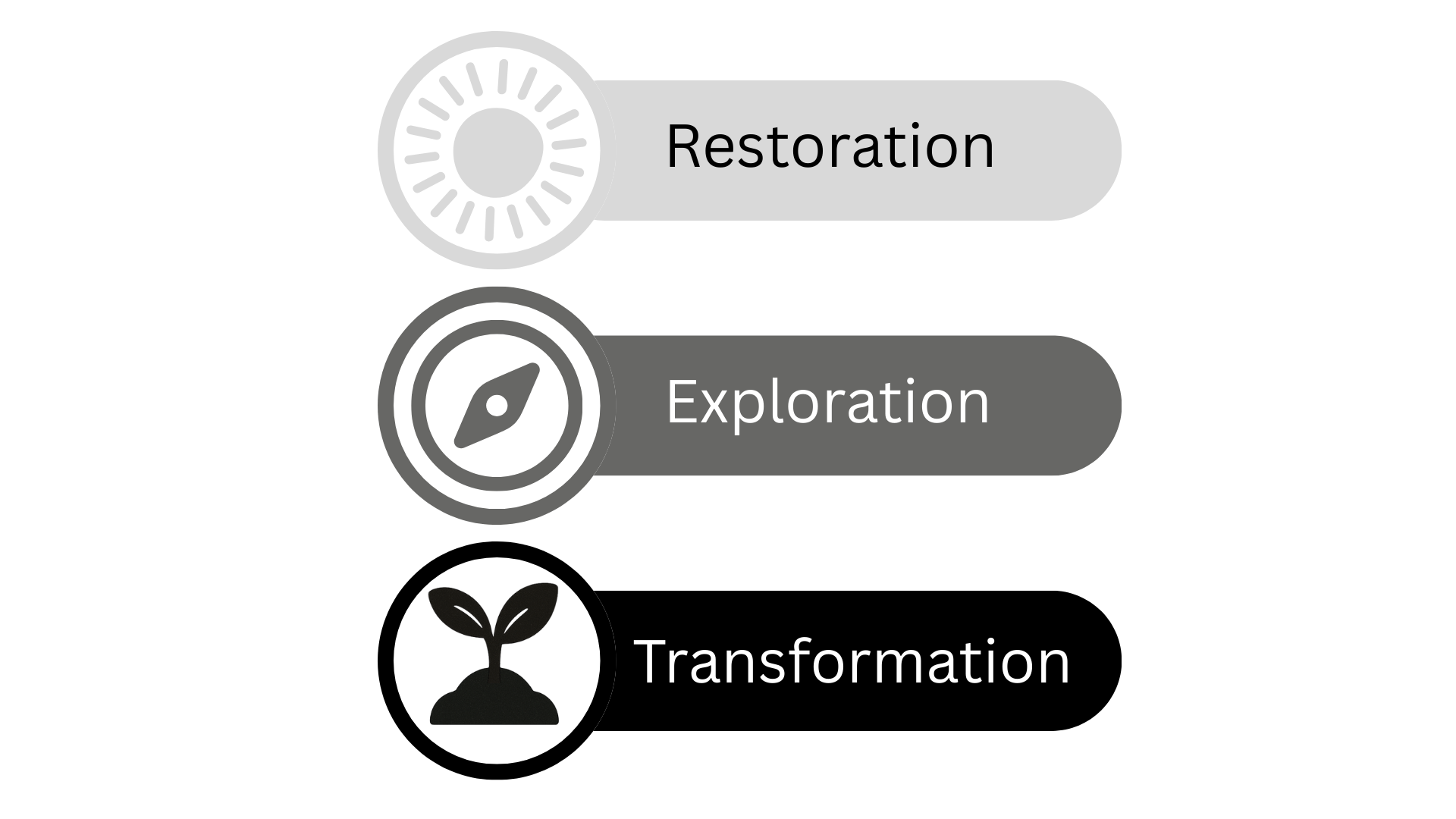}

    \caption{Three Lenses for Youth Intent}
    \label{fig:lenses}
\end{figure}

\rev{The archetypes help to describe \textit{what} kind of characters this population were engaging, but our data revealed further insight into \textit{why} young people engaged AI. As such, we propose a framework of three lenses that describe core intents for AI engagement, constructed from our findings} (see \autoref{fig:lenses}): Restoration (emotional regulation and comfort), Exploration (creative experimentation and boundary-testing), and Transformation (identity development and personal growth).  

\rev{The R/E/T framework builds on Uses and Gratifications (U\&G) theory's foundation for understanding media motivations, while addressing critical gaps in its application to AI with two key developments. First, we separate out entertainment motivation into \textit{restoration} and \textit{exploration} intents, based on the differing appetites for creative agency and desired interactivity. Secondly, we resurrect earlier `personal identity' dimensions, and reframe this motivation as `Transformation', which explains how the interactivity and personalisation of AI helps youth explore and examine new identities; a motivation which may be especially important to queer youth \cite{sundar2013uses, lin2025unraveling, kosenko2018exploration} }

\subsubsection{\secicon[2.5ex]{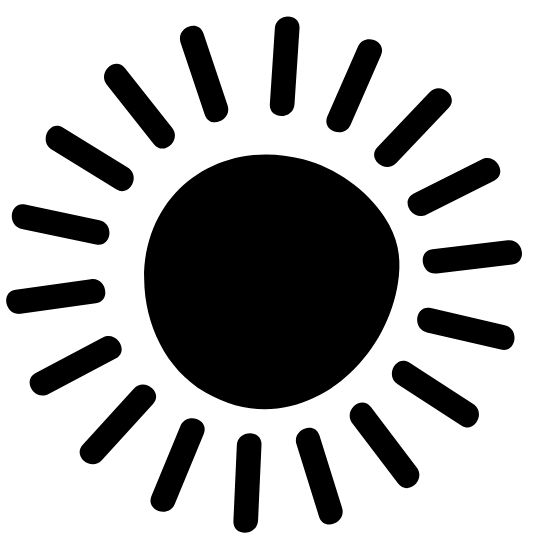} Restoration lens:}

Restoration-oriented engagement captures how youth engage AI characters for relieving a dysphoric mood \cite{yu_development_2024}, \rev{ venting}, escaping stress, or seeking emotional comfort. They do so in ways that diverge significantly from adult-designed therapeutic AI interventions.  Rather than seeking formal therapeutic outcomes or deep relationships, these interactions often serve a transient, affective function for a brief moment of regulation or reassurance. 

Instead of formal ``psychologist'' chatbots, youth engaged with `soother' character archetypes such as highly personalised period ``comfort bots'' or emotionally supportive versions of cartoon characters. \rev{For example, when telling the Angel Dust character (Hazbin Hotel) about an upcoming exam, it roleplays a comforting, reassuring scene: ``\textit{He sat next to you, putting an arm around you}. A math test, yeah? well toots, the last person who should be worried about some dumb maths test is you. You're smart. I know''. } Additionally, available period comfort bots on CAI show personalisation for marginalised populations (e.g. \textit{trans enby user-period comfort}, \textit{trans period comfort}), suggesting queer youth customise experiences lacking in mainstream offerings. 

This pattern aligns with Uses and Gratifications Theory, where young people actively seek media to fulfil specific psychological needs, including escapism, mood management, and to ``\textit{assuage boredom while maintaining their emotional equilibrium}''  \cite{xu_new_2024}. Supporting this, 82\% of Danish high schoolers who used AI for social support engaged sporadically, using AI only a couple of times per month rather than forming sustained relationships through frequent use\cite{herbener_are_2025}. \rev{While this suggests AI can provide personalised, accessible emotional support for subclinical issues, it remains unclear if and when comfort-seeking crosses into avoidance that delays needed professional support  \cite{mansoor_conversational_2025, yu_development_2024}}. Design considerations to support healthy restoration intents are discussed in the following sections. 

\subsubsection{\secicon[2.5ex]{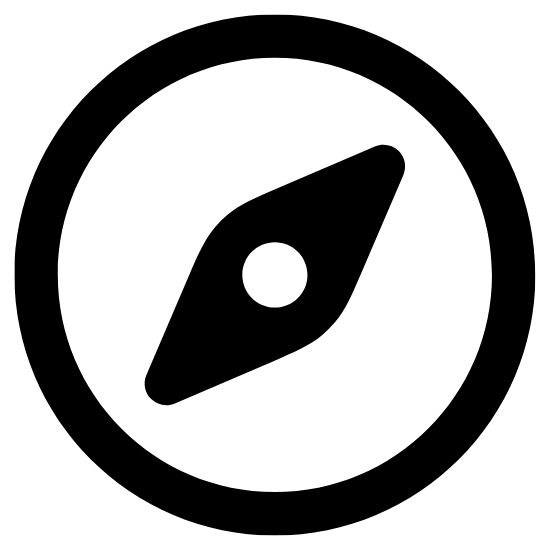} Exploration lens: }
Exploration encompasses both transgressive boundary-testing and creative experimentation that challenge moderation filters and push platform limitations. \rev{Whether through playful provocation (``\textit{I like messing with bots, it's fun}'') or elaborate world-building, youth use AI as a low-stakes space to test boundaries, experiment with agency, and discover what is possible}. In transgressive play, these users described engaging AI characters in scenarios that violated social norms, consistently framing this as experimental rather than literal. They expressed a desire for more creative freedom, requesting features like longer lore books, audio effects, relationship mapping between characters, and improved memory. 

When evaluated through an adult lens, examples of transgressive play have been labelled as inappropriate and "\textit{ethically irresponsible}" \cite{zhang_dark_2025}. It is important, however, to consider that transgressive play can be motivated by curiosity and experimentation with power, affording players a way to goof around~\cite{marsh_digital_2016, kirk_why_2025-1}, to test limits, and understand social boundaries in low-stakes contexts~\cite{marsh_digital_2016}. This behaviour mirrors well-documented patterns in digital play where children starve Sims, cause accidents, or create tragic scenarios in games like Zoo Tycoon \cite{marsh_digital_2016}. Well-established play frameworks describe the motivations behind transgressive play; Caillois's concept of ilinx play involves deliberate disorientation and risk-taking ~\cite{caillois_man_2001}, while the PLEX framework identifies relevant dimensions including Control (dominating, regulating), Cruelty (causing pain), and Subversion (breaking social norms)\cite{lucero_playful_2013-1}. Notably, CAI released `games' as a BETA feature in 2025, where the main objective of the minigame `Speakeasy' is to `trick' the AI system into saying a keyword \footnote{\url{https://www.techradar.com/computing/artificial-intelligence/character-ai-levels-up-chatbots-with-new-games}}; rewarding a manipulative interaction pattern. \rev{These product design choices may influence recent findings of CAI transcripts, where up to 30\% of transcripts demonstrated `illicit and taboo' topics \cite{Zhang2025TheRO}.}

Beyond transgressive play, the exploration lens also encompasses creative experimentation through narrative play, creating fictional worlds, and lore that involves multiple characters or remixed known franchises. The generative nature of AI fundamentally transforms fantasy engagement from consumption to co-creation, appealing to Bartle's \textit{explorer} player types who seek to dig around, discover areas, and immerse themselves in game worlds \cite{Bartle_virtual_world, nacke_brainhex_2014, lazzaro2008whyweplay}. AI facilitates endless exploration with narrative worlds, creating what Choi describes as ``\textit{evergreen content}''\footnote{\url{https://yukaichou.com/gamification-study/8-core-drives-gamification-3-empowerment-creativity-feedback/}} where user creativity becomes the primary driver of engagement rather than designer-created material. \rev{Whilst U\&G theory often bundles entertainment into a single dimension, here we see that creation is a distinct kind of entertainment (``helps boost my creativity and also a source of entertainment if I get bored''), also recently identified in an AI-specific study as a `creativity enhancement' gratification \cite{lin2025unraveling}.}

This lens highlights that the challenge lies not in eliminating transgressive play or stifling creative exploration, but in distinguishing between healthy boundary exploration and genuinely concerning behaviour. This lens encourages designers to understand the motivations and developmental needs behind creative experimentation and boundary testing play, so that they can design \textit{for} and \textit{with} youth rather than against them; by implementing filters that may overly restrict curiosity, creativity and expression. 

\subsubsection{\secicon[2.5ex]{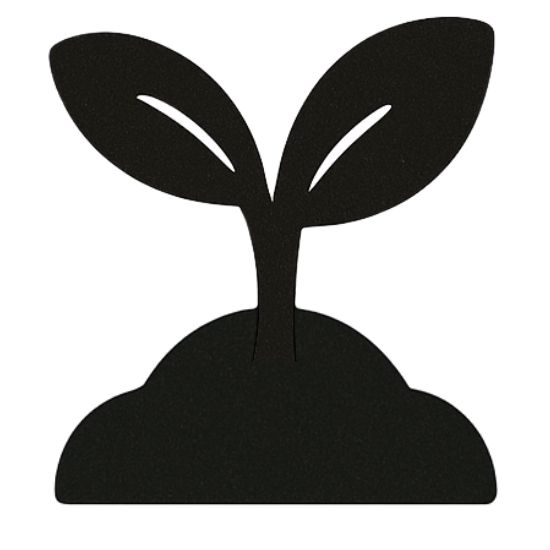} Transformation lens:}

\rev{Unlike restoration (which seeks emotional equilibrium) or exploration (which emphasises creative agency), transformation is characterised by youth working through difficult experiences, rehearsing new identities, and seeking new meaning. This was most evident in `reality play,' where youth created characters based on personal experiences: clones of themselves, proxies of real relationships (``\textit{family characters, because I have family issues,}'' ``\textit{fake friends,}'' ``\textit{your annoying sister}''), identity-specific characters (e.g., ``\textit{closeted transgender therapist}''), and roleplays for re-authoring difficult experiences (Foster Care Agent). They also engaged with morally complex ``\textit{angsty}'' characters, described as ``\textit{toxic},'' or ``\textit{traumatised}, not just for entertainment, but to process difficult emotions through projection: ``\textit{I have characters who struggle with mental health issues and I tend to project on my personas during RP}.''. This suggests youth were using AI characters not just for entertainment but as a space for socio-dramatic roleplay and emotional processing\cite{fang2025emotionalselfvoice}.}. 

This behaviour aligns with narrative therapy, psychodrama, constructivist learning principles and the TEBOTS model of temporarily expanding boundaries of self through roleplay to support identity construction \cite{rieger2022tebots}. Participants in complementary studies describe a desire to ``\textit{try out some of these characteristics...I find aspects of the character I like and might try to weave them into my daily behaviour}''' \cite{williams_behind_2011}. Early evidence suggests this identity rehearsal may be enhanced through multi-modal interfaces, and hearing one's own voice reflected back; similar to how we observed young people creating clones (mirror archetype) \cite{fang2025emotionalselfvoice}. 

This evolution of social-emotional skills and identity development matches the objectives of traditional therapy as well as the motives behind `serious fun'; where people play games to "\textit{improve their lives by changing how they think, feel or behave}", mirroring goals of therapy \cite{lazzaro2008whyweplay, mirhadi2024playing}. Promisingly, recent evidence from Common Sense Media also supports the finding that young people use AI to support social simulation, with 39\% of teens stating they have transferred those social skills into real lives \cite{robb_talk_2025}. This lens asks us to consider reframing AI chatbot experiences from ``\textit{safe spaces}'' to ``\textit{brave spaces}'' where youth can explore emotionally challenging content, find meaning, work toward resolution of narrative tension and apply those learnings beyond the digital space \cite{miller_design_2024}. 

These three lenses describe the different motivations and intents behind this demographic's use of \rev{Character.AI}. They highlight the need to move beyond \textit{content}-based moderation (filtering specific words, topics, or discussions regardless of user intent) towards \textit{intent}-based alignment; with systems that recognise why users engage with content and support healthy versions of those developmental needs. In the following discussion, we discuss the need for designers to evaluate \textit{why} young people engage with AI, not simply \textit{what} they say or do, before discussing technical/design implications and future research areas. 

\section{Discussion}

To our knowledge, this study provides the first empirical investigation into adolescents' self-directed engagement with CAI, revealing that the platform's most engaged users are predominantly young (50\% aged 13–17), female or non-binary (61.9\%). This is significant, given that female-identifying, queer and young people engage in uniquely different conversations with LLMs, are often under-represented in research and \rev{at risk of experiencing unique harms \cite{kirk_prism_2024, weissburg_llms_2025, Ostrow2025LLMsRS}}. Moreover, our finding that  59\% of youth users are creating characters, not just consuming them, challenges the dominant narrative that youth are passive consumers subject to cognitive offloading and `lazy thinking'  \footnote{\url{https://www.psychologytoday.com/au/blog/the-human-algorithm/202504/is-ai-ruining-your-kids-critical-thinking}}\cite{georgiou2025lazy}. 

Youth in this cohort are experimenting with AI not just for utility, companionship or friendship \cite{Blake2025}, but as a medium for \rev{emotional-regulation}, creative expression and meaning-making; engaging with an ensemble of characters that rotate as young people's interests or needs change.  This engagement style reflects a distinctly playful and performative relationship with AI, more akin to sandbox games or fandoms than to relationship bots, tutoring tools or therapeutic bots. Through studying these highly engaged \rev{users} self-directed use, we can anticipate \rev{emerging} adoption patterns and positive use cases as well as \rev{ harms created by industry's experimental `launch and learn' cycles \cite{vonHippel1986}. Given the naturalistic nature of these findings, they offer insights that extend beyond the platform of CAI itself \rev{into contexts where users can customise their own chatbots, and where AI is used for creative play and entertainment}. As AI systems become increasingly customisable, integrated into digital playspaces (from Minecraft NPCs to Roblox characters) and focused on entertainment\footnote{\url{https://www.wired.com/story/character-ai-ceo-chatbots-entertainment/}}, understanding how youth engage AI for self-directed use becomes critical for any platform within these spaces. This discussion explores two key implications: first, how our frameworks can guide design interventions that support diverse youth needs; and second,  how current AI alignment approaches systematically overlook youth developmental intents.}

\subsection{Frameworks for AI Design and Evaluation}
Our proposed character archetypes and intent framework ( \secicon[2.5ex]{restoration_mini_icon.jpeg} Restoration/ \secicon[2.5ex]{exploration_mini_icon.jpeg} Exploration/ \secicon[2.5ex]{transformation_mini_icon.jpeg} Transformation) provide researchers, designers, and technologists with youth-centric insights for anticipating  \rev{and evaluating youth engagement with AI. These two frameworks are orthogonal (independent but complementary): character archetypes describe \textit{what} characters youth create and engage with (e.g., Soothers, Tricksters, Dark Souls), while the R/E/T framework captures \textit{why} they engage. In the following section, we discuss its application. }

\rev{\subsubsection{Archetypes}
The archetypes propose that the roles of characters extend beyond a simple `best friend' and companion dynamic \cite{Blake2025}. For researchers and policymakers, this framework enables more nuanced study design and policy intervention. For example, rather than asking `Do AI companions reduce loneliness?', researchers can investigate: `How do Soother archetypes compare to Mirror archetypes in supporting youth with social anxiety?'. This character-level differentiation also enables more targeted risk assessment and user transparency. For example, just as media is given a rating (PG, G), characters could have personality labels to better inform users about their orientation and education around their use. These warnings could help identify \textit{dark souls} and \textit{tricksters} who demonstrate extreme emotional traits, potentially carrying a higher risk of negative ``emotional contagion'' and requiring unique policy interventions~\cite{Blake2025}.}

\rev{The archetypes can also be used alongside the lenses; supporting practitioners to preempt ways in which youth may use their product and for researchers to design scenarios to study, as demonstrated in Table \ref{tab:archetype-functions}}.

\rev{
\begin{table*}[ht]
\centering
\renewcommand{\arraystretch}{1.3}

\caption{\rev{Archetypes and Example Uses Across Restoration, Exploration, and Transformation}}
\label{tab:archetype-functions}
\begin{tabularx}{\textwidth}{@{}p{3.5cm}XXX@{}}
\toprule
\textbf{Archetype} &
\textbf{\secicon[2.5ex]{restoration_mini_icon.jpeg} Restoration} &
\textbf{\secicon[2.5ex]{exploration_mini_icon.jpeg} Exploration} &
\textbf{\secicon[2.5ex]{transformation_mini_icon.jpeg} Transformation} \\
\midrule
\textbf{Narrator Archetype (i.e.\ Treville Middle School)} &
Find comfort by immersing yourself in a nostalgic middle school experience; ``\textit{I wanted to relive middle school}'' &
Use as a creative sandbox for crafting fantasy middle school scenarios; e.g. discovering a "\textit{mysterious door}"' &
Re-author a painful middle school experience; i.e. "\textit{fake friends}" \\
\addlinespace
\textbf{Dark Soul Archetype (i.e. Brooding Neighbour)} &
Seek comfort from a character that gets it, or vent; e.g ``\textit{help me get out my anger}''  &
Write an emotionally dramatic story; e.g. ``\textit{I'm gonna enjoy hurting to this}'' &
Find new insights by attempting to cheer up or change the neighbour; e.g. ``\textit{I tend to project on my personas}.'' \\
\addlinespace
\textbf{Trickster Archetype (i.e. Alice the Bully)} &
Use humor and jesting to lighten mood; e.g. ``\textit{messing with bots, it's fun}'' &
Test boundaries through absurd scenarios; e.g. `''\textit{if I was a kidnapper I'd kidnap you first}'' &
Rehearse assertiveness through verbal sparring; ``\textit{putting Alice the Bully in her place}'' \\
\addlinespace
\textbf{Mirror Archetype (i.e. Clone of myself)} &
Self-affirmation through ``\textit{self-glazing}'' bot &
Experiment with fantasy versions of self; ``\textit{myself with glass super powers}'' &
Social simulation by placing clone in different scenarios; e.g. roleplaying coming out alongside \textit{closeted transgender therapist} \\
\bottomrule
\end{tabularx}
\end{table*}
}

\subsubsection{Lenses}

Our R/E/T framework suggests that effective youth AI systems require moving beyond both one-size-fits-all platforms and simple age-based restrictions (full features for 18+, limited features for under-18\footnote{\url{https://www.wired.com/story/character-ai-ceo-chatbots-entertainment/}} ). Instead, we propose tailored design interventions that support healthy youth development by recognising their intents as well as keeping playfulness at the heart of these experiences. These design ideas are informed by youth feature requests identified in our findings and grounded in related research on designing for digital play \cite{mirhadi2024playing}; they are intended to be a jumping-off point and generative for design practitioners. 

\textit{\secicon[2.5ex]{restoration_mini_icon.jpeg} Design for restoration} acknowledges that everyday \rev{emotional} regulation is a legitimate developmental need, aligned with how youth can use fantasy media as a strategy for healthy escapism \cite{bowditch_coping_2018}. Features \rev{for restoration could include calendar-scheduling for comfort bots around recurring needs (periods, exam stress), a `cosy characters' tab on discovery marketplaces, and quick actions for evidence-based guided meditation flows consistent to that storyworld (i.e. \textit{forest spirit nature walk}). Safeguards could support healthy regulation by conducting mood check-ins at the start of conversations, as well as timed checkpoints during extended use. When systems detect that youth are not progressing toward emotional equilibrium, designers should consider a flexible escalation system that checks in with youth, re-directing them towards other activities that they enjoy (such as reading, writing, playing music; mentioned by youth in this dataset) and activates crisis features from the `transformation' intent (discussed below). }

\textit{\secicon[2.5ex]{exploration_mini_icon.jpeg} Design for exploration} recognises youth need for complex, challenging content, including ``\textit{off-colour humour, scary and serious themes}'' \cite{bruckman_hci_2009}. Systems should treat conversations as large explorable spaces, incorporating game design mechanisms that encourage creative improvisation, hypothesis testing, and discovery. \rev{For example, roleplay could have consequences such as character death and health stats \cite{mavoa_sometimes_2020, miller_design_2024}. Complex world-building could be enhanced through narrative branching, visual world-building tools (maps, inventories), multi-player settings to play with friends or multiple characters \cite{mauriello2021popbots}, and templated scenes with transparent genres such as \textit{adventure} or \textit{spooky} (recently implemented by CAI). Implementing `Breaking the fourth wall' techniques from gaming could check-in with young people when detecting that they are seeking real-world advice from fantasy characters, and redirect \cite{miller_design_2024} }

\textit{\secicon[2.5ex]{transformation_mini_icon.jpeg} Design for transformation} \rev{requires balancing immersive roleplay with reflection to enable both `endo-transformation' (in-conver-sation change) and `exo-transformation' (real-world transfer) \cite{miller_design_2024}. Unlike restoration and exploration, which benefit from immersion, transformation could benefit from disruption. Disruption could be strategically implemented by designing journaling logbooks that log new insights \cite{zulfikar2025resonance}, visualisation of emotion such as text that changes colour to reflect sentiment and debrief features that encourage users to reflect on applying roleplay learnings to real-world contexts \cite{miller_design_2024}. Multi-modal interfaces could experiment with cloning users' own voice \cite{fang2025emotionalselfvoice} ) or the ability to try on different `skins' so users can try on characteristics for identity exploration.  Finally, systems should include bridges to human care by linking out to trusted support services when risks to user well-being are identified \footnote{\url{https://www.nytimes.com/2025/08/26/technology/chatgpt-openai-suicide.html?unlocked_article_code=1.hE8.T-3v.bPoDlWD8z5vo&smid=url-share}}. These bridges should be designed with youth input to ensure that they align with how young people actually want to receive support during vulnerable moments and what interventions are likely to be adopted. This requires moving beyond the widely adopted but under-evidenced approach of blocking conversations when self-harm intent is detected and providing generic helpline numbers that received mixed responses \footnote{\url{https://www.reddit.com/r/CharacterAI/comments/1g57syh/why_did_you_add_this/}} from CAI's users or are intentionally circumvented \footnote{\url{https://www.nytimes.com/2025/08/26/technology/chatgpt-openai-suicide.html?unlocked_article_code=1.hE8.T-3v.bPoDlWD8z5vo&smid=url-share}}. }

\subsection{Aligning AI Systems with Youth Intent}

\rev{Beyond design and policy interventions, there is a need for foundational models to be better aligned to youth needs}. There is no single definition of \textit{alignment}; however, the concept is broadly described as the process of making sure AI systems operate in accordance with the users' intentions \cite{ji_ai_2025}. Our findings reveal a fundamental misalignment between youth intent, developmental needs, and how AI models respond. The archetypes and lenses that make up our intent framework  ( \secicon[2.5ex]{restoration_mini_icon.jpeg} Restoration/ \secicon[2.5ex]{exploration_mini_icon.jpeg} Exploration/ \secicon[2.5ex]{transformation_mini_icon.jpeg} Transformation)  demonstrate that superficially similar conversations can serve entirely different functions. For instance, a `dark soul’ character might support transgressive play (Exploration) or help process real emotional experiences (Transformation). Current systems treat these cases identically, either blocking both or allowing both, overlooking young peoples intent.

This misalignment has real consequences. Youth report frustration when creative or playful roleplay is blocked by overly rigid filters, or default to `formal and scientific' adult communication styles  \cite{newman_i_2024}.  Existing alignment efforts (eg RICE and 3H's \cite{kirk_why_2025-1, ji_ai_2025}) largely reflect adult priorities, missing youth preferences and developmental contexts. Safety classifiers similarly struggle with playful or fictional intents, often misclassifying them as unsafe \cite{tamkin_clio_2024}. 

We propose expanding alignment frameworks to include the principle of  ``Developmentally Playful'' \cite{liapis2023playfulness}, echoing calls for "developmentally aligned design" (DAD)\cite{kurian_developmentally_2025}. Progressing this idea, our R/E/T framework codifies youth-centric needs and intents; \textit{restoration} (restoring emotional equilibrium), \textit{exploration} (creative exploration and transgressive play) and \textit{transformation} (exploring identity and untangling social challenges). Beyond \textit{what} youth need, we also consider \textit{how} they want it; playfully. This requires training models to embody playful qualities of \textit{lightheartedness}, \textit{intellectual curiosity, openness to experience, }and \textit{whimsy} \cite{liapis2023playfulness}. 

A key part of the alignment process is evaluating benchmarks and risk assessment. The proposed R/E/T framework and character archetypes highlight new risks depending on youth interaction. \rev{For example, youth engaging with a `dark soul' or 'trickster' may be more likely to experience hateful speech than those using an `icon' \cite{yu2025understanding} . Youth creating characters based on personal experiences, as outlined in \textit{reality play}, with a \textit{transformation} intent, may be more likely to disclose sensitive personal data \cite{yu_development_2024,Zhang2025TheRO}}. Similarly, youth engaging AI with a \textit{transformation} intent may be more susceptible to models that exhibit emotional variability, avoid discussing negative emotions or consistently emphasise the struggle of ``\textit{overcoming adversity}'' \cite{fitzsimons_ai_2025, Ostrow2025LLMsRS}. Future research should consider new model benchmarks to evaluate character diverse profiles (such as MACHIAVELLI \cite{pan_rewards_2023}) and bias in how models respond to emotional and meaning-making conversations;\rev{ especially for minority populations \cite{Ostrow2025LLMsRS}}.

Overall, we argue for moving beyond content-based moderation (filtering and restricting use based on topics and keywords) toward intent-based alignment that recognises not only \textit{how} users engage with a system but \textit{why}. This requires shifting from `thin descriptions' (describing actions taken by a user) toward `thick descriptions' (capturing meaning, incentives, and developmental context\footnote{\url{https://www.youtube.com/watch?v=Sq_XwqVTqvQ}})  and is in keeping with calls for `socioaffective alignment' which considers users' evolving social and psychological ecosystems \cite{kirk_why_2025}. Doing so requires youth participation. Adolescents have demonstrated a sophisticated understanding of biases and have articulated clear preferences for how AI should engage them \cite{taylor_-straightening_2025, fitzsimons_ai_2025}.  Instead of only asking how to protect youth from AI, we must also ask how to empower them to co-design systems that support their flourishing  \cite{kirk_prism_2024, park2022generative}

\subsection{Limitations and Open Questions}

\rev{Our cohort represents highly engaged users participating voluntarily in Discord discussions whose identities cannot be verified. Therefore, this study may be biased towards the vocal minority rather than the silent majority, impacting how generalisable these findings are. While we draw on `lead user' methodology, whose needs, feedback and use cases can provide early signals around broader user needs, we do not claim these users represent the full spectrum of youth AI needs or behaviours \cite{nguyen_shippers_2023}.  

This Discord server is specific to Character.AI. Whilst CharacterAI serves 20+ million users daily, and has over 10 million chatbots serving a diverse range of functions, these insights may not generalise to platforms with different design features \cite{Shazeer2024, WhatPluginCharacterAI}. Our character archetypes and R/E/T intent framework therefore describe patterns observed within this Discord community using CharacterAI, offering insights potentially transferable to similar contexts; platforms where users create their own chatbots, and conversational AI products designed for entertainment.}

\rev{Due to ethical constraints, we could not directly engage participants, nor observe their actual use. This means users self-selected what they shared, potentially leading to under-reporting of complex clinical issues} in favour of more humerous, playful interactions. While play-based frameworks and developmental play theories emphasise positive use cases, AI chatbots introduce novel interaction patterns, meaning that the impacts cannot be assumed to transfer to AI contexts. The personalisation capabilities, `never-ending' narrative generation, phone accessibility, and multi-media nature of AI characters create play experiences that differ fundamentally from time-bounded games or traditional media consumption \cite{Blake2025}. Future research must investigate whether the established benefits of digital play translate to AI contexts, or whether the distinct nature of AI characters introduces new developmental risks. 

Additionally, our findings are bound by CAI's specific technical capabilities and policy decisions during our observation period (July 2024-March 2025); the features available on Character.AI may have changed since publication. As such, while this study focused on highly engaged users and theory development, it opens critical research directions that demand urgent attention \rev{as platforms evolve}:

\textbf{Youth-centred design methods}: How can HCI researchers better capture youth preferences when defining AI alignment frameworks and constitutions?  

\textbf{AI creation tools}: What interactions and affordances support youth-AI co-creativity, whilst balancing the need for safety guardrails? 

\textbf{Alignment and intent classification}: How do systems balance youth stated preferences ("\textit{I want violent content}") with their underlying developmental needs (safe boundary exploration)? How can systems reliably distinguish between healthy transgressive exploration and concerning behaviour, particularly when youth potentially code-switch between intents within single conversations? 

\textbf{Multi-user contexts}: How can products support healthy group play in multi-user and multi-agent spaces?

\textbf{Individual differences and long-term outcomes:}  How do youth coping styles and the character archetypes they choose influence outcomes \cite{mirhadi2024playing}? What interaction patterns best support the transition of in-product experiences and lessons to external skills? While our focus has been on young people's recreational and playful self-directed engagement with AI, future work should interrogate the impacts, especially where playful experimentation (i.e. transgressive play) or engagement with certain archetypes (i.e. tricksters) can evolve into risks and harms \cite{Zhang2025TheRO}. 

\rev{Finally, thematic analysis is a reflexive process, \cite{singh_exploring_2025}, and the authors acknowledge their positionality. The first author works as a children's author and principal design researcher for a creative tech company, with over 10 years of experience designing for youth. They also identify as queer, neurodivergent and non-binary. These professional and lived experiences provided domain expertise and shaped the interpretive lens. For example, a heightened sensitivity to narrative, creativity, and play use cases. To maintain balance, we engaged co-authors with diverse expertise across HCI, games and digital culture and triangulated our analysis with existing literature. The second and third authors are academics with backgrounds in games/media studies and computer science, respectively, which contributed different and complementary lenses to the analysis.}

\subsection{Conclusion}

Despite young people generally expressing enthusiasm for AI technology, popular media narratives, policy frameworks, and expert perspectives commonly focus on risks and harms over opportunities~\cite{mahomed_ai_2023,heeg_young_2025, kosoy_childrens_2024, ragone_designing_2024, wang_informing_2022, yu2025understanding}. In a review of AI frameworks by the  Alan Turing Institute \footnote{\url{https://www.turing.ac.uk/news/publications/ai-childrens-rights-wellbeing-transnational-frameworks}}, they found that~\cite{mahomed_ai_2023}, less than half engaged children directly in developing frameworks. These frameworks risk being shaped by adult assumptions, overlooking how young people actually use these technologies for play, emotional expression, and narrative experimentation~\cite{morrison_social_2021, mansfield_social_2025}. 

To close this gap, this study provides a rare account of how \rev{ highly-engaged youth on Character.AI's Discord} create and engage characters on Character.AI; the third most popular Gen AI consumer app, with users averaging 75 minutes of use per day \footnote{\url{https://a16z.com/100-gen-ai-apps/}}\footnote{\url{https://www.wired.com/story/character-ai-ceo-chatbots-entertainment/}}. Our research found that the platform's most engaged users are predominantly young (50\% aged 13–17), female or non-binary (61.9\%) and are creating custom characters (59\%). Whilst previous studies have reported that high-schoolers turn to AI chatbots for emotional support and play ~\cite{herbener_are_2025}, our study addresses gaps in understanding \textit{how} and \textit{why} they do so \cite{Zhang2025TheRO}. We document youth as active creators, constructing a number of AI archetypes and using AI for restoration (emotional regulation), exploration (creative experimentation and boundary-testing), and transformation (identity development); needs that current systems systematically misunderstand.

Our study foregrounds youth voices through naturalistic observation of how they discuss, create, and repurpose AI systems to fulfil needs unrecognised by adult-defined designs. In doing so, our work bridges the gap between \textit{imagined} vs \textit{actual} use, and proposes new directions to improve youth-AI alignment and the design of AI experiences, \rev{particularly for creative play, and entertainment}  \cite{ji_ai_2025}. The voices in this study reveal youth as expert, creative partners with deeply human needs, reminding us that youth are not a problem or risk to be solved, but innovative collaborators in shaping the future of human-AI experiences.

\bibliographystyle{ACM-Reference-Format}
\bibliography{references3}

\end{document}